\documentclass[aps,pre,showkey,square,numbers,amssymb,amsmath,longbibliography,nobibnotes]{revtex4-1}
\usepackage{graphicx,color}
\usepackage{bm}% bold math
\usepackage{epsfig}
\usepackage{verbatim}
\usepackage{hyperref}

\providecommand{\red}[1]{\textcolor{black}{#1}}

\providecommand{\blue}[1]{\textcolor{black}{#1}}

\begin{document}

%Title of paper
\title{Cost of excursions until first crossing of the origin for random walks
and L\'evy flights:\\ an exact general formula}
\author{Francesco Mori$^1$, Satya N. Majumdar$^2$, Pierpaolo Vivo$^3$}
\affiliation{$^1$ Rudolf Peierls Centre for Theoretical Physics, University of Oxford, Oxford, United Kingdom\\$^2$ LPTMS, CNRS, Univ. Paris-Sud, Universit\'e Paris-Saclay, 91405 Orsay, France\\$^3$ Department of Mathematics, King’s College London, London WC2R 2LS, United Kingdom}

\begin{abstract}

We consider a discrete-time random walk \blue{on a line} starting at $x_0\geq 0$ where a 
cost is incurred at each jump.  We obtain an exact analytical formula 
for the distribution of the total cost of a trajectory until the process 
crosses the origin for the first time. The formula is valid for 
arbitrary jump distribution and cost function (heavy- and light-tailed 
alike), provided they are symmetric and continuous. We analyze the 
formula in different scaling regimes, and find a high degree of 
universality with respect to the details of the jump distribution and 
the cost function. Applications are given to the motion of an active 
\blue{run-and-tumble} particle in one dimension and extensions to multiple cost 
variables are 
considered. The analytical results are in perfect agreement with 
numerical simulations.

\end{abstract}

\maketitle

\tableofcontents

\section{Introduction}

In a variety of disciplines, key events occur when a stochastic process 
reaches a predefined target state for the first time 
\cite{Redner_book,BF_2005,sm13,fp_book_2014}. For instance, in finance, limit 
orders are employed to execute trades when a specified price level is 
reached. Similarly, in the study of foraging behavior among animals 
\cite{foraging} it is important to estimate the typical duration of an 
exploration phase, until the animal goes back to its home for the first 
time.

\blue{These problems can be thoroughly characterized using the 
theory of first passage events of stochastic processes}. 
Consider the paradigmatic case of a 
discrete-time random walk \blue{on a line where the position
$x_m$ of the walker at step $m$ evolves via}
\begin{equation}
    x_m= x_{m-1}+ \eta_m\,,
    \label{markov.1}
\end{equation}
\blue{starting from} $x_0\geq 0$ and 
where the jumps $\eta_m$'s are independently 
drawn from a symmetric and continuous \blue{probability \red{density} 
function (PDF)} 
$f(\eta)$. A central quantity to characterize first-passage events 
is the probability $F_n(x_0)$ that the walker crosses the origin 
for the first time at step $n$, \blue{starting from $x_0\ge 0$}. 
A cornerstone of first-passage theory is the 
\emph{Sparre Andersen theorem} \cite{SA_54}, 
which states that when the walker starts from the origin ($x_0=0)$, 
the first-passage probability 
\red{
\begin{equation}
   F_n(0)= \frac{(2(n-1))!}{n!(n-1)!}2^{-2n+1} \label{eq:fn0_intro}
\end{equation}
}
is completely universal, i.e., 
independent of the jump distribution $f(\eta)$, as long 
as $f(\eta)$ is continuous and symmetric. Crucially, the universal 
behavior is valid for any finite $n$ and even for L\'evy flights, 
corresponding to fat-tailed $f(\eta)\sim 1/|\eta|^{\mu+1}$, with 
$0\!<\!\mu<2$. This universality extends to \blue{many other
observables for the random walk}, e.g., the 
number of records \blue{up to step $n$}
\cite{MZ_2008, Louven_review, MSW_2012, GMS16, GMS_2017,CM_2020}, 
and \blue{also to other processes via mapping to random walks.
Examples of such processes include
continuous-time run-and-tumble particles  
\cite{MLMS_20a,MLMS_20b} and 
resetting processes in one dimension \cite{KMSS_2014,MMSS_2022,SMS23} }.

However, the classical treatment of the \blue{first-passage} problems 
usually ignores an important variable: 
there are often costs or rewards, e.g., in terms of monetary fees or 
energy consumption, associated with the change of state of the process. 
For instance, in the example of animal foraging, 
there is an \emph{energy cost} associated to a roaming trajectory -- 
as well as a potential \emph{energy gain} if food is actually found 
along the way. \blue{In such contexts,} 
\textit{Markov reward models} \cite{reward1,reward2,reward3}, 
where a Markov process drives an auxiliary cost/reward dynamics 
have proven useful in \blue{various fields such as} 
wireless communications \cite{wireless1,wireless2}, 
biochemistry \cite{biochemistry}, insurance models \cite{insurance} and 
software development \cite{software}. To describe the cost 
associated to the process, \blue{one couples} to the random walk in 
Eq.~\eqref{markov.1} a cost variable $C_n$, evolving according to
\begin{equation}
C_m= C_{m-1} + h(\eta_m)\,,
\label{cost.1_intro}
\end{equation}
with $C_0=0$. The cost function $h(\eta)>0$ is assumed to be continuous and
symmetric, i.e., $h(\eta)=h(-\eta)$, but is otherwise arbitrary.
The function $h(\eta)$ can be interpreted as the energy spent or 
the cost incurred
by the walker in making a single jump $\eta$. 
Cost dynamics of the type \eqref{cost.1_intro} with a non-linear 
function $h(\eta)$ have been used to model the fare structure of 
taxi rides, static friction under random applied forces, and depinning 
transition in spatially inhomogeneous media 
\cite{MMV23.1,MMV23.2,old1,old2,old3,old4,old5,old6,old7}. 
Additionally, recent works have investigated the cost associated to 
resetting processes \cite{FGS16,MSK23,SBEM,PPPL_24,SBEM_24}. 

\begin{figure}
\includegraphics[width=0.6\textwidth]{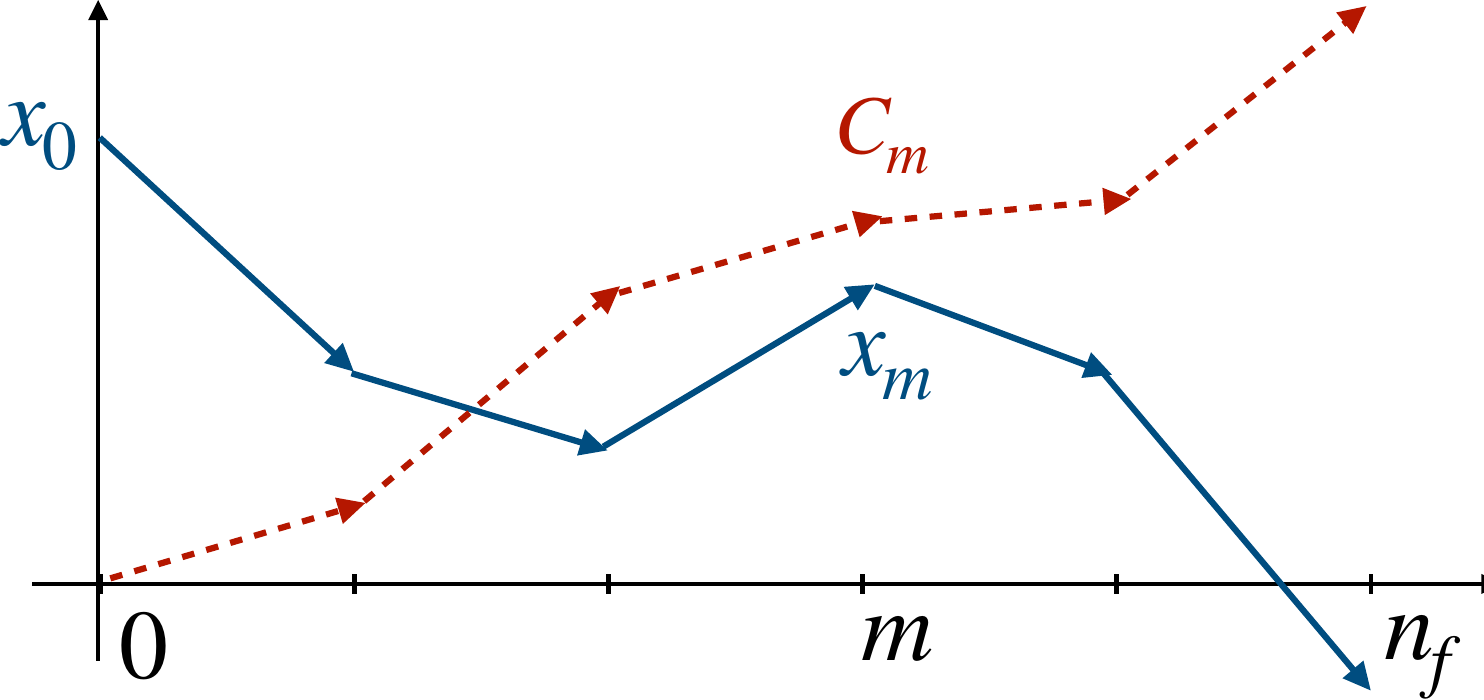}
\caption{Typical trajectory of a random walk $x_m$ on a line starting at 
$x_0\ge 0$ and evolving
in discrete time via the jump process \eqref{markov.1}. The process 
stops at step $n_f$ when the walker crosses the origin for
the first time. A cost $C_m$ is associated with the trajectory of the 
random walk up to step $m$.}
\label{fig.rw1}
\end{figure}

We focus here on first-passage processes of the random walk variable 
$x_m$. The process stops when the walker, starting from $x_0\geq 0$, 
crosses the origin for the first time at step $n_f$ (see 
Fig.~\ref{fig.rw1}). Note that $n_f$ itself is a random variable that 
fluctuates from one trajectory to another. Given $f(\eta)$ and 
$h(\eta)$, we are interested in computing the distribution $Q(x_0,C)$ of 
the total cost $C=\sum_{m=1}^{n_f} h(\eta_m)$ till the first-passage 
time to the origin, \blue{given that the walker 
starts its journey at $x_0\ge 0$. Note that, 
for symmetric and continuous $f(\eta)$ and $h(\eta)$,
by shifting the origin to $x_0$ and reversing the time, one finds
that $Q(x_0,C)$ also gives the cost distribution of a random
walk, starting at the origin, till the first-hitting time
of the level $x_0$.}

As a \blue{concrete} and paradigmatic application, the cost distribution 
till the first-passage time
appears \red{naturally} in the run-and-tumble particle (RTP) model
of an active particle in one dimension. 
Active particles consume energy directly from the environment
and move via self-propelled motions. For example, {\em E. Coli} bacteria
typically move in space by alternating phases of `run' and `tumble'
\cite{Berg_book,TC_2008}.
Consider for simplicity the standard RTP model in one dimension 
\cite{TC_2008,ALP_2014,Malakar_2018}.
The particle starts at \blue{some position $x_0\ge 0$, \red{it} chooses a velocity $\pm v_0$ with equal probability}
and runs ballistically with this velocity during
a random time $\tau$ chosen from the exponential distribution
$p(\tau)= \gamma\, e^{-\gamma\, \tau}$ where $1/\gamma$ is the
persistence time. After this initial run, the particle
`tumbles' instantaneously, i.e., it chooses a new velocity
$\pm v_0$ with equal probability and then moves
ballistically again with this chosen velocity during
another random period drawn from $p(\tau)$.
Then it tumbles again and the dynamics continues via
the alternating `run' and `tumble' phases.
\blue{Now let us imagine that a target, such as a food \red{item},
is located at the origin. The RTP stops when it reaches the target,
i.e., it goes past the origin.}
A natural question then is how much energy does the RTP spend, \blue{starting
its motion at $x_0\ge 0$}, till
\blue{it finds the target?} This model can be clearly
mapped to the Markov jump process in Eq.~\eqref{markov.1}
where $x_n$ denotes the position of the RTP after the $n$-th
run -- this is also known as a `persistent random walk'~\cite{Kac_74}.
This mapping has been recently used successfully to derive
interesting universal results for the survival probability
of a generalized RTP model with arbitrary velocity distribution
after each tumbling~\cite{MLMS_20a,MLMS_20b}.
Via this mapping, the run length $\ell_n$ of the
$n$-th run in the standard RTP model (with $\pm v_0$ velocities)
is precisely the jump length $\eta_n$ of the random walk
model in Eq.~\eqref{markov.1}. Since, $\ell_n= v_0\, \tau_n$
where $\tau_n$ is the exponentially distributed run time of the $n$-th run,
it follows that this RTP model then corresponds precisely to the jump distribution
\begin{equation}
f(\eta)= \frac{\gamma}{2\, v_0}\, e^{- \frac{\gamma}{v_0}\, |\eta|}\, ,
\label{exp_jump.1}
\end{equation}
in the random walk model in Eq.~\eqref{markov.1}.
One can then associate a cost function $h(\eta)$ denoting
the energy spent during each run. \blue{Consequently,
the quantity $Q(x_0,C)$ is precisely the distribution 
of the cost incurred, or the energy consumed by the RTP
till it finds its target, given the fixed initial starting
position $x_0\ge 0$. For continuous and symmetric $f(\eta)$
and $h(\eta)$, as before, $Q(x_0,C)$ also gives the
distribution of the cost for an RTP starting at the origin
till it finds its target located at $x_0\ge 0$.} \red{ Note that, as clarified in Section \ref{Model}, for simplicity we only interrupt the RTP trajectory when the first tumbling event occurs after the target is found.}

Beyond its natural interpretation as a cost function, the variable $C$ 
can describe a broad class of \emph{first-passage functionals} of 
discrete-time random walks. Indeed, \blue{ by choosing different
$h(\eta)$ appropriately, one can obtain a whole class of $C$'s
describing different observables associated to a first-passage 
trajectory.} For instance: (i) with $h(\eta)=1$, the variable $C$ simply 
coincides with the first passage time $n_f$ itself, (ii) with 
$h(\eta)=|\eta|$, the 
variable $C$ describes the total distance traveled by the walker, and (iii) with 
$h(\eta)=\theta(|\eta|-\eta_c)$, where $\theta(z)$ is the Heaviside step 
function, $C$ describes the number of steps longer than $\eta_c$. In the 
continuous-time setting, first-passage functionals of stochastic 
processes have been widely studied in the literature with many 
applications \cite{MB02,MC05,MM20,SP22,M23,R23} (for a review see 
Ref.~\cite{BF_2005}). Note, however, that most of these works have 
considered functionals depending on the \emph{state} of the process 
$x_m$, instead of the step size $\eta_m$, i.e., of the form 
$C=\sum_{m=1}^{n_f} h(x_m)$. On the other hand, for discrete-time random 
walks and functionals depending on the step size, there are, to the best 
of our knowledge, much fewer general results.
\blue{A notable exception to this
is} a series of recent works \cite{ACEK14,BCP22,Pozzoli_phd_thesis}
where, using a general combinatorial theory, the cost distribution $Q(C)\equiv Q(0,C)$ 
was computed for the case $x_0=0$.  
\blue{Thus $Q(C)$ is the distribution of the cost of the trajectory till
its first return to the starting point, i.e., the origin.}
In particular, in Ref.~\cite{BCP22} a formula for 
the Fourier transform of $Q(C)$ was derived, valid for arbitrary step 
distribution $f(\eta)$ and cost function $h(\eta)$. 
%This general result provides an exact relation between the 
%Fourier transform of $Q(C)$ and quantities that do not depend on 
%the first-passage events, which are in general simpler to derive. 
This formula allows to derive analytically the distribution $Q(C)$ for a 
variety of random-walk models.

\blue{In this work, our goal is to generalize this result for $x_0=0$ 
to the cost distribution $Q(x_0,C)$ for arbitrary starting position $x_0\ge 0$.
Alternatively, for a walker starting at the origin and with
symmetric and continuous $f(\eta)$ and $h(\eta)$,
our result for $Q(x_0,C)$ provides the cost distribution till the 
first-hitting time of the level $x_0\ge 0$, thus generalizing the
result for the cost distribution till the first-return time to the
origin. 
Our main result is to derive, by solving exactly an integral equation, 
an explicit general formula for 
the double Laplace transform of $Q(x_0,C)$ in terms of $f(\eta)$ and $h(\eta)$, valid for continuous and
symmetric $f(\eta)$ and $h(\eta)\ge 0$. This exact formula reads
\begin{equation}
{\hat Q}(\lambda,p)= \int_0^{\infty}dx_0\int_0^{\infty}dC\,
e^{-p\, C}\, e^{-\lambda\, x_0}\, Q(x_0,C)
=\frac{1}{\lambda}\,
\left[ 1- \sqrt{1- 2\, A(p)} \, \psi_1 (\lambda, p)\right]\, ,
\label{sol.2}
\end{equation}
where $A(p)= \int_0^{\infty} f(\eta)\, e^{-p\, h(\eta)}\, d\eta$ and
\begin{equation}
\psi_1(\lambda, p)= \exp\left[- \frac{\lambda}{\pi}\, \int_0^{\infty}
\frac{ \ln \left(1-\, \int_{-\infty}^{\infty} 
f(\eta)\,e^{-p\, h(\eta)}\, e^{i\, q\, \eta}\,
d\eta\right)}{\lambda^2+q^2}\, dq\right]\ .
\label{psi_def.1}
\end{equation}
For $x_0=0$, our formula recovers the result of Refs.~\cite{ACEK14,BCP22}
obtained by a completely different combinatorial method. As mentioned above,
our main interest in this paper is on $Q(x_0,C)$ for general $x_0\ge 0$,
going beyond $x_0=0$ and to explore, for example, if any universal behavior emerges
for large $x_0$.
While our formula for ${\hat Q}(\lambda,p)$,
stated in Eqs. (\ref{sol.2}) and (\ref{psi_def.1}), 
is exact, inversion of this double Laplace
transform leading to an explicit $Q(x_0,C)$ is very hard
for general $f(\eta)$ and $h(\eta)$. In fact, even extracting
the asymptotic behaviors of $Q(x_0,C)$ for different limiting
values of $x_0$ and $C$ is highly nontrivial. Nevertheless,
exploiting this exact formula, we were able to extract the 
limiting scaling behaviors of $Q(x_0,C)$
in different regimes of $x_0$ and $C$, for different choices of
$f(\eta)$ and $h(\eta)$.
We provide a
detailed analysis of different scaling regimes involving $C$ and $x_0$,
for both light- and heavy-tailed jump distributions, as well as several
instances of specific distributions for which the calculations can be
performed explicitly and compare very well with numerical simulations.
Our results bring out the hidden universalities of the cost distribution
as a function of $x_0$. As an additional bonus, we also
provide several explicit exact results for specific choices of
$f(\eta)$ and $h(\eta)$ in the case $x_0=0$, that were not
studied in \cite{ACEK14,BCP22} and which can be easily verified
numerically.} Our results also extend 
naturally to several cost variables, with interesting 
applications to prey-predators models \cite{prey} and excursions in 
environments with feedback-coupling \cite{feedback}.

\blue{The plan for the rest of the paper is as follows. 
In Section \ref{Model}, we first present a more general
random walk-cost model and derive an exact formula for
the Fourier-Laplace transform of $Q(x_0,C)$ (see Eqs. (\ref{sol.1})
and (\ref{psi_def})). In Section \ref{additive},
we focus on the additive cost function model and derive
our main results for $Q(x_0,C)$ in Eqs. (\ref{sol.2})
and (\ref{psi_def.1}). In Section \ref{general_x0}, 
we analyze the distribution $Q(x_0,C)$ in different regimes of $x_0$
and investigate its asymptotic 
behaviors when the mean cost and the step variance are finite. 
In Section \ref{section:levy}, we investigate the cost 
associated to L\'evy flights with L\'evy index $\mu$ in the case 
$h(\eta)=|\eta|$, so that the cost $C$ describes the total length of 
the trajectory until the first-passage events. 
We derive the different regimes of $Q(x_0,C)$ and derive 
universal scaling functions, considering the cases 
$1<\mu<2$ and $0<\mu<1$ separately. 
In Section \ref{sec:x0zero}, for $x_0=0$, we consider several models 
for which $Q(0,C)$ can be computed exactly and compared
directly to numerical simulations. Finally, 
in Section \ref{sec:conclusions} we conclude and present some future 
directions.}

\section{Distribution of Cost for a random walk/L\'evy flight till its first-passage time: general theory}
\label{Model}

\blue{We begin with a more general setting where the position $x_m$ of
the random walker on a line at step $m$, starting at $x_0$, evolves again 
by the Markov rule
\begin{equation}
x_m= x_{m-1}+ \eta_m\ ,
\end{equation}
where the jumps $\eta_m$'s are independent and identically distributed (IID).
Following Refs.~\cite{ACEK14,BCP22,Pozzoli_phd_thesis}, 
we consider an 
associated cost function
that evolves via 
\begin{equation}
C_m= C_{m-1} + \xi_m\, \quad {\rm starting}\,\, {\rm from}\,\, C_0=0\, ,
\label{cost.1}
\end{equation}
where $\xi_m$'s are also IID random variables.
At any given step $m$, the jump increment $\eta_m$ and the cost
increment $\xi_m$ may \red{be} in general correlated and we assume
that the pair $(\eta_m,\xi_m)$
is drawn from a \red{joint density} $p(\eta,\xi)$. We also
assume that the walk is symmetric, i.e., $p(\eta,\xi)=p(-\eta,\xi)$. 
Note that the case of additive costs, introduced in Eq. (\ref{cost.1_intro})
and on which we will focus in this work, can be recovered by 
setting $p(\eta,\xi)=f(\eta)\delta(\xi-h(\eta))$. But below
we will first derive a general result valid for arbitrary
$p(\eta,\xi)=p(-\eta,\xi)$ and then focus on the special additive cost
case later.}

\blue{We are interested in the probability distribution $Q(x_0,C)$ of the \red{total} cost 
$C$ at the
time when the position $x_m$ crosses the origin for the first time, given
that the walk starts at $x_0\ge 0$. Note
that, for the moment, the cost \blue{$C_m$ at step $m$} can be both 
positive and negative.}
Then $Q(x_0,C)$ satisfies the backward equation
\begin{equation}
Q(x_0,C)= \int_0^{\infty} dx' \int_{-\infty}^{\infty} dC' Q(x', C') p(x'-x_0, C-C')
+ \int_{-\infty}^{-x_0} p(\eta_1,C)\, d\eta_1\, ,
\label{bfp.1}
\end{equation}
where $x'$ is the position after the first jump. 
The two terms on the right hand side (RHS) of Eq.~\eqref{bfp.1} reflect two 
distinct cases. The first term corresponds to the case where the walker 
starting from $x_0$ jumps to a new position $x'>0$ with a cost $C-C'$ 
associated to this first jump. The remaining cost until first passage 
time must therefore be $C'$. The second term describes trajectories in 
which the walker crosses the origin at the first step \blue{to a new
position $x'<0$, i.e., the first jump $\eta_1<-x_0$}, with an 
associated cost $C$. \red{Note that, for simplicity, we include in the total cost $C$ the full cost $h(\eta_{n_f})$ associated with the last step $\eta_{n_f}$. In other words, we do not interrupt the random walk trajectory at $x=0$, but we let the walker complete its last step, so that $x_{n_f}<0$.}

We now define the Fourier transform
with respect to $C$
\begin{equation}
\tilde{Q}(x_0,k)= \int_{-\infty}^{\infty} dC\, e^{i\, k\, C}\, Q(x_0, C)\, .
\label{FT.1}
\end{equation}
\blue{Note that the Fourier transform, as opposed to the Laplace transform, is an
appropriate choice here since $C$ can be both positive and negative at this stage.} 
Taking Fourier transform of Eq.~(\ref{bfp.1}) with respect to $C$ gives
\begin{equation}
\tilde{Q}(x_0,k)= \int_0^{\infty} dx'\, \tilde{Q}(x', k)\, 
\int_{-\infty}^{\infty} p(x'-x_0, \xi)\, e^{i\,k\, \xi}\, d\xi + 
\int_{-\infty}^{-x_0} \left[ \int_{-\infty}^{\infty} p(\eta_1,\xi)\, 
e^{i\, k\, \xi}\, d\xi\right]\, d\eta_1\, .
\label{FT_bfp.1}
\end{equation} 
Let us define a renormalized jump probability density function (PDF), parametrized by $k$,
\begin{equation}
f_k(\eta)= \frac{\int_{-\infty}^{\infty} p(\eta,\xi)\, 
e^{i\, k\, \xi}\, d\xi}{\int_{-\infty}^{\infty} d\eta\, 
\int_{-\infty}^{\infty} d\xi\, p(\eta,\xi)\, 
e^{i\, k\, \xi}}\, ,
\label{jump_renorm.1}
\end{equation}
which is symmetric in $\eta$, continuous and is normalized to unity. 
Let us also denote the denominator in Eq.~(\ref{jump_renorm.1}) by
\begin{equation}
g(k)= \int_{-\infty}^{\infty} d\eta\, 
\int_{-\infty}^{\infty} d\xi\, p(\eta,\xi)\, e^{i\, k\, \xi}\, .
\label{gk_def}
\end{equation}
Hence, in terms of $f_k(\eta)$ and $g(k)$, Eq.~(\ref{FT_bfp.1}) can be expressed as
\begin{equation}
\tilde{Q}(x_0,k)= g(k)\, \int_0^{\infty} dx'\, \tilde{Q}(x', k)\, f_k(x'-x_0) +
g(k)\, \int_{\red{-\infty}}^{-x_0} f_k(\eta_1)\, d\eta_1\, .
\label{FT_bfp.2}
\end{equation}
\blue{Our next goal is to solve this integral equation, which
is of the general Wiener-Hopf variety (where the integral limit
in the first term on the RHS starts at $0$ indicating a
semi-infinite domain). Such Wiener-Hopf integral
equations are notoriously difficult to 
solve~\cite{Louven_review}, except for few special cases (see Appendix \ref{appendix} for a special case where an explicit solution can be obtained). 
Fortunately, as we show below, our integral equation (\ref{FT_bfp.2})
can be mapped to one of these solvable cases.} 

\blue{To see how we can solve Eq. (\ref{FT_bfp.2}), 
let us first consider a different random walk process $x_m=x_{m-1}+\zeta_m$, 
starting at $x_0$,
where $\zeta_m$'s are IID variables}, each drawn from, say, $p(\zeta)$
\blue{which is assumed to be symmetric and continuous, but otherwise arbitrary}. 
Let $F(x_0,n)$ denote the probability that starting at $x_0$, the walker crosses the
origin for the first time at step $n$. Then $F(x_0,n)$ satisfies the 
backward equation
\begin{equation}
F(x_0,n) = \int_0^{\infty} F(x', n-1)\, p(x'-x_0)\, dx' + \delta_{n,1}\, 
\int_{-\infty}^{-x_0}
p(\zeta_1)\, d\zeta_1 \, .
\label{fp.1}
\end{equation}
\blue{The first term on the RHS describes what happens
after the walker jumps from $x_0$ to a new value $x'\ge 0$ at the first step.
This happens with probability density $p(x'-x_0)$ and following the jump,
the process, \red{starting} now at $x'$, needs to cross the origin for the 
first time at step $(n-1)$--this occurs with probability $F(x',n-1)$.
Multiplying the two and integrating over all $x'\ge 0$ gives the
first term on the RHS. The second term on the RHS describes the
event when the walker crosses the origin by the first jump itself and
lands on a position $x'<0$, i.e., the first jump $\zeta_1<-x_0$.}  

Defining the generating function
\begin{equation}
\tilde{F}(x_0,s)= \sum_{n=1}^{\infty} F(x_0,n)\, s^n\, ,
\label{genf.1}
\end{equation}
one finds from Eq.~(\ref{fp.1}) that the generating function satisfies the integral equation
\begin{equation}
\tilde{F}(x_0,s)= s\, \int_0^{\infty} \tilde{F}(x',s)\, p(x'-x_0)\, dx' + 
s\, \int_{-\infty}^{-x_0}
p(\zeta_1)\, d\zeta_1\, .
\label{fp.2}
\end{equation}
Fortunately this integral equation can be explicitly solved and
its solution is given by the Pollaczek-Spitzer 
formula~\cite{Pollaczek_52,Spitzer_56,Spitzer_57} (for a pedagogical account
see ~\cite{Louven_review}), which states that the Laplace transform of $\tilde{F}(x_0,s)$
with respect to $x_0$ is given by
\begin{equation}
\int_0^{\infty} dx_0\, e^{-\lambda\, x_0}\, \tilde{F}(x_0,s)= \frac{1}{\lambda}\left[1-\sqrt{1-s}\, 
\phi(\lambda, s)\right]\, ,
\label{PS.1}
\end{equation}
with 
\begin{equation}
\phi(\lambda, s)= \exp\left[- \frac{\lambda}{\pi}\, \int_0^{\infty} \frac{\ln \left(1-s\, 
\hat{p}(q)\right)}{\lambda^2+q^2}\, dq\right]\, .
\label{phi_def}
\end{equation}
Here $\hat{p}(q)$ is the Fourier transform of the jump distribution
\begin{equation}
\hat{p}(q)= \int_{-\infty}^{\infty} d\zeta\, e^{i\, q\, \zeta}\, p(\zeta)\, .
\label{FTpq.1}
\end{equation}

We now note the formal similarity between the integral equation (\ref{FT_bfp.2}) and
the one in (\ref{fp.2}) with the identifications
\begin{equation}
s\equiv g(k) \quad {\rm and}\quad p(\eta)\equiv f_k(\eta)\, .
\label{equiv.1}
\end{equation}
Consequently, we can solve (\ref{FT_bfp.2}) using the Pollaczek-Spitzer solution of (\ref{fp.2})
and the identifications in Eq.~(\ref{equiv.1}). Hence, our exact solution of Eq.~(\ref{FT_bfp.2})
can be stated in terms of the Laplace-Fourier transform of the cost distribution, namely,
\begin{equation}
\int_0^{\infty} \tilde{Q}(x_0,k)\, e^{-\lambda \, x_0}\, dx_0= \frac{1}{\lambda}\,
\left[ 1- \sqrt{1- g(k)} \, \psi (\lambda, k)\right]\, ,
\label{sol.1}
\end{equation}
with 
\begin{equation}
\psi(\lambda, k)= \exp\left[- \frac{\lambda}{\pi}\, \int_0^{\infty}
\frac{ \ln \left(1- g(k)\, \int_{-\infty}^{\infty} f_k(\eta)\, e^{i\, q\, \eta}\, 
d\eta\right)}{\lambda^2+q^2}\, dq\right]\, ,
\label{psi_def}
\end{equation}
where $f_k(\eta)$ is the renormalized jump PDF defined in Eq.~(\ref{jump_renorm.1}) and
$g(k)$ is given in Eq.~(\ref{gk_def}).
Eqs.~(\ref{sol.1}) and (\ref{psi_def}) are our main results valid for 
arbitrary $x_0$ and \blue{$p(\eta,\xi)$}
and below we will discuss several interesting special cases. 

The first corollary of our general result concerns the limit 
$x_0\to 0$, i.e., when the walk starts from the
origin. Indeed, making the change of variable $\lambda\, x_0=u$ on 
the left hand side (LHS)
of Eq.~(\ref{sol.1}) and taking the $\lambda\to \infty$ limit on both sides of (\ref{sol.1})
gives
\begin{equation}
\tilde{Q}(0,k)= \int_{-\infty}^{\infty} Q(0,C)\, e^{i\, k\, C}\, dC 
= 1- \sqrt{1- g(k)}= 1- \sqrt{1- \int_{-\infty}^{\infty} d\eta\,
\int_{-\infty}^{\infty} d\xi\, p(\eta,\xi)\, e^{i\, k\, \xi}}\, .
\label{Q0k.1}
\end{equation}
This is exactly the result derived in 
Refs.~\cite{ACEK14,BCP22,Pozzoli_phd_thesis} by generalizing a combinatorial method
that was originally used to prove the Sparre Andersen theorem~\cite{SA_54}.
Sparre Andersen theorem has many applications in physics and 
probability theory, for
reviews see~\cite{Feller,Redner_book, Persistence_review,fp_book_2014,Louven_review}.
\blue{Our method involving the solution of the integral equation (\ref{bfp.1}),
thus provides an alternative derivation of the cost distribution
in the special case ($x_0=0$).}

\subsection{Additive cost function}
\label{additive}

\blue{As discussed in the introduction, our main interest in this paper
is to analyze $Q(x_0,C)$ for general $x_0\ge 0$, for the case
of additive cost functions where $\xi_m= h(\eta_m)$
with $h(-\eta)=h(\eta)>0$. This case has many applications
as discussed in the introduction.} 
In this case, the joint \red{density} of $\eta$ and $\xi$
is given by 
\begin{equation}
p(\eta, \xi) = f(\eta)\, \delta(\xi-h(\eta))\, .
\label{jpdf_noise.1}
\end{equation}
\blue{Hence, the Fourier transform \red{\eqref{gk_def}} simplifies to}
\begin{equation}
g(k)= \int_{-\infty}^{\infty} d\eta\, f(\eta)\, e^{i\, k\, h(\eta)} \, ,
\label{gk_sp.1}
\end{equation}
and
\begin{equation}
f_k(\eta)= \frac{f(\eta)\, e^{i\, k\, h(\eta)}}{g(k)}\, .
\label{fk_sp.1}
\end{equation} 
Also, since $h(\eta)$ is a symmetric function, the cost $C\ge 0$, 
and hence, it is convenient to 
define the Laplace transform 
(instead of the Fourier transform) by making the analytic continuation 
$k=i\, p$. Then we have,
using the symmetry of $f(\eta)$ and $h(\eta)$,
\begin{equation}
g(k) \to 2\, A(p) \quad {\rm where} \quad A(p)= \int_0^{\infty} f(\eta)\, e^{-p\, h(\eta)}\, d\eta\, ,
\label{Ap_def}
\end{equation}
and
\begin{equation}
f_k(\eta)\to f(\eta,p)= \frac{f(\eta) e^{-p h(\eta)}}{2\, A(p)}\, .
\label{fep.1}
\end{equation}
Consequently, our general result in Eqs.~(\ref{sol.1})
and (\ref{psi_def}) gives the result for the additive cost function 
stated in Eqs. (\ref{sol.2}) and (\ref{psi_def.1}) which we re-state
here for later convenience
\begin{equation}
\int_0^{\infty} dx_0\, \int_0^{\infty} dC\,  Q(x_0,C)\,e^{-p\, C}\, e^{-\lambda \, x_0}= \frac{1}{\lambda}\,
\left[ 1- \sqrt{1- 2\, A(p)} \, \psi_1 (\lambda, p)\right]\, ,
\label{sol.22}
\end{equation}
where $A(p)=\int_0^{\infty} f(\eta)\, e^{-p\, h(\eta)}\, d\eta$ and
\begin{equation}
\psi_1(\lambda, p)= \exp\left[- \frac{\lambda}{\pi}\, \int_0^{\infty}
\frac{ \ln \left(1-\, \int_{-\infty}^{\infty} f(\eta)\,e^{-p\, h(\eta)}\, e^{i\, q\, \eta}\,
d\eta\right)}{\lambda^2+q^2}\, dq\right]\, .
\label{psi_def.11}
\end{equation}
Once again, for $x_0=0$, the computation simplifies
(following the $\lambda\to \infty$ limit) and we get
\begin{equation}
\int_0^{\infty} dC\, Q(0,C)\, e^{-p\, C}=1-\sqrt{1- 2\, A(p)}\ .
\label{sol_sp_0}
\end{equation}
\blue{Later in Section \ref{sec:x0zero},
we will show that Eq. (\ref{sol_sp_0})
can be explicitly inverted in
several cases, which can subsequently be compared easily to numerical
simulations.}

\section{The dependence of $Q(x_0,C)$ on the starting position $x_0$}
\label{general_x0}
%\subsection{Case $x_0>0$ (finite mean cost per jump and 
%variance of jump distribution)}

Our main interest here is to investigate how the cost distribution $Q(x_0,C)$ depends on the
distance $x_0$ between the initial position of the searcher and the target. We will consider
here the case when the cost increment $\xi_n=h(\eta_n)$, where $h(\eta)$ is a symmetric and continuous
function of $\eta$. In this case, the cost $C$ is always positive and the starting point of 
our analysis is the exact formula stated 
in Eqs.~(\ref{sol.2}) and (\ref{psi_def.1}). While this general formula is valid for
arbitrary jump PDF $f(\eta)$ (as long as $f(\eta)$ is symmetric and continuous), in order
to keep things simple, we will first focus in this section on the case when 
the mean cost per jump and the variance of the jump 
distribution are both finite and are given respectively by 
\begin{equation}
\mu_1= \int_{-\infty}^{\infty} h(\eta)\, f(\eta)\, d\eta \quad {\rm and}\quad 
\sigma^2= \int_{-\infty}^{\infty} \eta^2\, f(\eta)\, d\eta \, .
\label{mean_jump_def}
\end{equation} 
In this case, it turns out that $Q(x_0,C)$ for large $C$, as a function of $x_0$, has two 
principal
regimes: (1) when $x_0\sim O(1)$  and (2) when $x_0\sim O\left(\sqrt{C}\right)$. 
These two regimes also emerged in the analysis of the survival probability
of the random walker in Eq.~(\ref{markov.1})  up to $n$ steps, starting at $x_0$~\cite{MMS_17}.
Our analysis below will proceed along similar lines as in Ref.~\cite{MMS_17} and below
we present the analysis for the two regimes
separately. 

\subsection{When $x_0\sim O(1)$ for large $C$}
\label{Sec:finite.1}

To analyze this regime, we need to take the limit $p\to 0$ with $\lambda$ fixed in
Eqs.~(\ref{sol.2}) and (\ref{psi_def.1}). To leading order for small $p$, we get
\begin{equation}
2\, A(p)\simeq 1- \mu_1\, p \, .
\label{Ap_small.1}
\end{equation}
Plugging this expansion into Eq.~(\ref{sol.2})
and setting $p=0$ in $\psi_1(\lambda, p)$ in (\ref{psi_def.1}) we find
\begin{equation}
\int_0^{\infty} dx_0\,  e^{-\lambda \, x_0}\, \int_0^{\infty} dC\,  Q(x_0,C)\,e^{-p\, C}\simeq
\frac{1}{\lambda}\, \left[1- \sqrt{\mu_1\, p}\,\, \psi_1(\lambda,0)\right]\, ,
\label{x1.1}
\end{equation}
where
\begin{equation}
\psi_1(\lambda,0)= \exp\left[-\frac{\lambda}{\pi}\, 
\int_0^{\infty} \frac{\ln (1- \hat{f}(q))}{\lambda^2+q^2}\, dq\right]\, \quad 
{\rm with}\quad \hat{f}(q)= \int_{-\infty}^{\infty} f(\eta)\, e^{i\, q\, \eta}\, d\eta\, .
\label{x1.2}
\end{equation}
Taking derivative of Eq.~(\ref{x1.1}) with respect to $p$ gives
\begin{equation}
\int_0^{\infty} dx_0\,  e^{-\lambda \, x_0}\, \int_0^{\infty} dC\,  C\, Q(x_0,C)\,e^{-p\, C}\simeq
\frac{1}{\lambda}\, \frac{\sqrt{\mu_1}}{ \sqrt{4\,p}}\, \psi_1(\lambda,0)\, .
\label{x1.3}
\end{equation}
Inverting the Laplace transform with respect to $p$, using 
${\cal L}_{p\to C}^{-1}\left[p^{-1/2}\right]= 1/\sqrt{\pi C}$, we get
\begin{equation}
Q(x_0,C)\simeq \sqrt{\frac{\mu_1}{4}}\, \frac{1}{C^{3/2}}\, U_2(x_0)\, ,
\label{Qx0_scale.1}
\end{equation}
where the Laplace transform of the function $U_2(x_0)$ is given by
\begin{equation}
\int_0^{\infty} U_2(x_0)\, e^{-\lambda\, x_0}\, dx_0= \frac{1}{\lambda\, \sqrt{\pi}}\, 
\psi_1(\lambda,0)\, ,
\label{Ux0_def}
\end{equation}
with $\psi_1(\lambda,0)$ defined in Eq.~(\ref{x1.2}). Let us remark that exactly the
same function $U_2(x_0)$ appeared in the analysis of the survival probability
of the random walk (\ref{markov.1}) up to step $n$ and its asymptotic behaviors
were analyzed in great detail in Ref.~\cite{MMS_17}. Without repeating this analysis,
let us just summarize the asymptotic behaviors of $U_2(x_0)$~\cite{MMS_17}. One gets
\begin{eqnarray}
\label{Ux0_asymp}
U_2(x_0)=\begin{cases}
& \frac{1}{\sqrt{\pi}}+ \alpha_1\, x_0 + O(x_0^2) \quad\quad\quad\quad {\rm as}\quad x_0\to 0 \\
\\
& \frac{\sqrt{2}}{\sigma\, \sqrt{\pi}}\, \left(x_0+C_2\right) + O\left(\frac{1}{x_0}\right)
\quad\,\,\, {\rm as} \quad x_0\to \infty\, .
\end{cases}
\end{eqnarray}   
The constants $\alpha_1$ and $C_2$ are given explicitly by
\begin{equation}
\alpha_1= \frac{1}{\pi^{3/2}}\, \int_0^{\infty} dq\, \ln \left( 1- \hat{f}(q)\right)\, ;
\quad\quad {\rm and}\quad\quad C_2= -\frac{1}{\pi}\, \int_0^{\infty} \frac{dq}{q^2}\,
\ln \left[ \frac{1-\hat{f}(q)}{\sigma^2 q^2/2}\right]\, .
\label{constat_def}
\end{equation}
The constant $C_2$ was first computed analytically in Ref.~\cite{CM_2005} as
the subleading correction term in the asymptotic growth of the expected maximum
of a random walk as a function of the number of steps. Interestingly, the
same constant also appeared in the context of an efficient rectangle packing algorithm in 
computer science~\cite{Coffman_98}, in the mean perimeter of the convex hull of 
a two dimensional random walk~\cite{GLM_17}, in the Smoluchowsky trapping problem 
for Rayleigh flights in three dimensions~\cite{MCZ_06,ZMC_07,Ziff_91}, and as the so 
called `Milne constant' or `Hopf constant'
in the radiative transfer of photons~\cite{Milne_1921,PS_1947,Finch}.

We end this subsection by noting that exactly for $x_0=0$, using $U(0)=1/\sqrt{\pi}$ from
Eq.~(\ref{Ux0_asymp}), we get from (\ref{Qx0_scale.1})
\begin{equation}
Q(0,C)\simeq \sqrt{\frac{\mu_1}{4\, \pi}}\, \frac{1}{C^{3/2}}\, .
\label{Q0C.1}
\end{equation}
We will derive again this asymptotic result in Section \ref{subsec:asymptotic} (see Eq.~\eqref{QC_tail.1}) by taking $x_0=0$ first and then taking the limit of large $C$.

\subsection{When $x_0\sim \sqrt{C}$ for large $C$}

\begin{figure}
\includegraphics[width=0.45\textwidth]{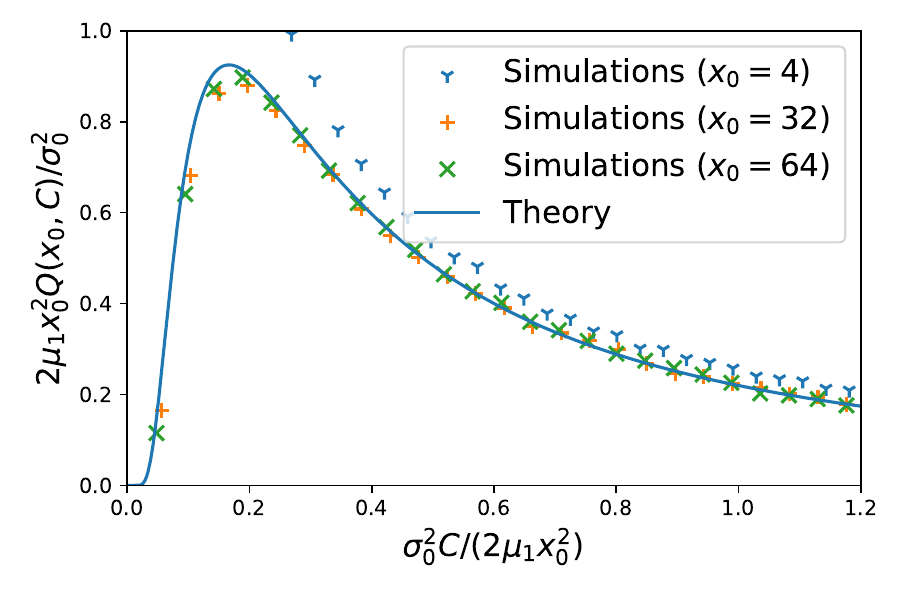}
\includegraphics[width=0.45\textwidth]{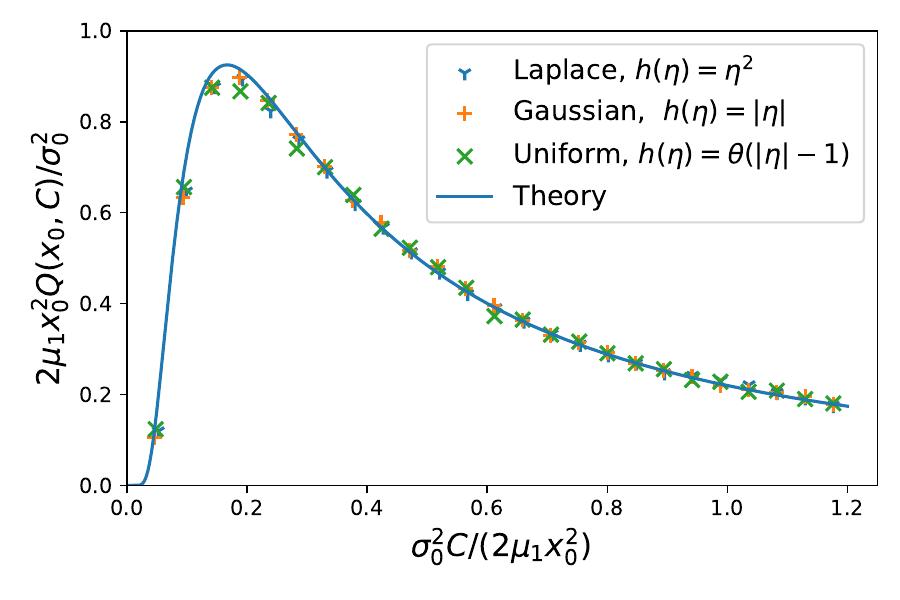}
\caption{Scaled probability density function 
$2\mu_1 x_0^2 Q(x_0,C)/\sigma_0^2$ as a function of the 
scaled cost $\sigma_0^2 C/(2\mu_1 x_0^2)$. The continuous blue line 
corresponds to the scaling form in Eq.~\eqref{Hz_scaling.1}, while 
symbols are obtained from numerical simulations, averaging over $10^6$ 
repetitions. The left panel shows how the numerical results collapse 
onto the theoretical one for increasing $x_0$, with 
$f(\eta)=e^{-|\eta|}/2$ and $h(\eta)=|\eta|$. The right panel is 
obtained from numerical simulations with $x_0=40$ and different choices 
of $f(\eta)$ and $h(\eta)$: $f(\eta)=e^{-|\eta|}/2$ (Laplace) with 
$h(\eta)=\eta^2$, $f(\eta)=e^{-\eta^2/2}/\sqrt{2\pi}$ (Gaussian) with 
$h(\eta)=|\eta|$, and $f(\eta)=\theta(2-|\eta|)$ (Uniform) with 
$h(\eta)=\theta(|\eta|-1)$, where $\theta(z)$ is the Heaviside step 
function. \blue{The fact that they all collapse onto the same scaling 
function demonstrates the universality with respect to the jump distribution
$f(\eta)$ with a finite variance.}}
\label{fig:collapse}
\end{figure}

In this case we consider the scaling when $C\to \infty$ and $x_0 \to \infty$, with the ratio
$C/x_0^2$ held fixed. Our main result here is to show that in this scaling limit, 
$Q(x_0,C)$ exhibits the
scaling form
\begin{equation}
Q(x_0,C) \simeq \frac{\sigma^2}{2 \, \mu_1\, x_0^2}\, H\left( \frac{\sigma^2\, C}{2\, \mu_1\, x_0^2}\right)\, ;
\quad {\rm where}\quad H(z)= \frac{1}{\sqrt{4\, \pi\, z^3}}\, e^{-1/(4z)}\quad {\rm for}\quad z\ge 0\, .
\label{Hz_scaling.1}
\end{equation}
The scaling function $H(z)$ is universal, i.e., it does not depend on the details
of $h(\eta)$ and $f(\eta)$ as long as $\mu_1$ and $\sigma^2$ are both finite (see Fig.~\ref{fig:collapse}).
For large $z$, the scaling function $H(z) \sim z^{-3/2}$ has a power law tail, while
it has an essential singularity as $z\to 0$. 

To extract the universal scaling form (\ref{Hz_scaling.1}) from our exact
results in Eqs.~(\ref{sol.2}) and (\ref{psi_def.1}) turns out to be somewhat
nontrivial as we now show. Since both $x_0$ and $C$ are large with their ratio
$C/x_0^2$ held fixed, we need to analyze  Eqs.~(\ref{sol.2}) and (\ref{psi_def.1})
in the scaling limit $\lambda\to 0$, $p\to 0$ with the ratio $\lambda/\sqrt{p}=w$
held fixed. Let us first analyze the function $\psi_1(\lambda,p)$ in Eq.~(\ref{psi_def.1})
in this scaling limit. To proceed, we first rescale $q=\lambda u$ in (\ref{psi_def.1})
which gives
\begin{equation}
\psi_1(\lambda, p)= \exp\left[- \frac{1}{\pi}\, \int_0^{\infty}
\frac{ \ln \left(1-\, \int_{-\infty}^{\infty} f(\eta)\,e^{-p\, h(\eta)}\, e^{i\, \lambda\, u\, 
\eta}\,
d\eta\right)}{1+u^2}\, du\right]\, .
\label{psi_def.2}
\end{equation}
Expanding the argument of the logarithm in small $\lambda$ and small
$p$, and using the definitions (\ref{mean_jump_def}), it is easy to see that, to leading order
for small $\lambda$ and small $p$ with $w= \lambda/\sqrt{p}$ held fixed, one gets
\begin{equation}
1-\, \int_{-\infty}^{\infty} f(\eta)\,e^{-p\, h(\eta)}\, e^{i\, \lambda\, u\, 
\eta}\,
d\eta \simeq p\, \left(\mu_1+ \frac{\sigma^2 w^2}{2}\, u^2\right)\, .
\label{scale.1}
\end{equation}
Substituting this in Eq.~(\ref{psi_def.2}) we can now perform the integral over $u$
explicitly using the identity
\begin{equation}
\int_0^{\infty} \frac{\ln (a+b\, u^2)}{1+u^2}\, du= \pi\, \ln\left(\sqrt{a}+ \sqrt{b}\right)\, ,
\quad {\rm for}\quad a>0\, ,\, b>0\, .
\label{identity.1}
\end{equation}
This gives
\begin{equation}
\psi_1(\lambda, p)\simeq 
\frac{1}{\sqrt{p}\, \left(\sqrt{\mu_1} + \frac{\sigma\, w}{\sqrt{2}}\right)}\, .
\label{psi_def.3}
\end{equation} 
Substituting Eq.~(\ref{psi_def.3}) and the small-$p$ expansion of $A(p)$ in Eq.~(\ref{Ap_small.1}) into (\ref{sol.2})
\begin{equation}
\int_0^{\infty} dx_0\, \int_0^{\infty} dC\,  Q(x_0,C)\,e^{-p\, C}\, e^{-\lambda \, x_0}\simeq
\frac{1}{\lambda}\, \frac{w}{ \left(w+ \frac{\sqrt{2\,\mu_1}}{\sigma}\right)}\, .
\label{dlt_scale.1}
\end{equation}
We then substitute the scaling form (\ref{Hz_scaling.1}) on the LHS
of Eq.~(\ref{dlt_scale.1}) and rescale $\lambda x_0= y$. Upon cancelling the factor
$1/\lambda$ from both sides of (\ref{dlt_scale.1}) and using $\lambda=w\, \sqrt{p}$ , we then get
\begin{equation}
\int_0^{\infty} dy\, e^{-y}\, \int_0^{\infty} dz\, H(z)\, 
e^{- 2\,\mu_1\, y^2\, z/(w^2\, {\sigma}^2)}
= \frac{w}{ \left(w+ \frac{\sqrt{2\,\mu_1}}{\sigma}\right)}\, .
\label{dlt_scale.2}
\end{equation}

Defining further $s= 2\mu_1/(\sigma^2 w^2)=s$, we get the following double Laplace transform
of the scaling function $H(z)$
\begin{equation}
\int_0^{\infty} dy\, e^{-y}\, \int_0^{\infty} dz\, H(z)\, 
e^{-s\, y^2\, z}= \frac{1}{1+\sqrt{s}}\, .
\label{dlt_scale.3}
\end{equation}
Let us emphasize that this relation (\ref{dlt_scale.3}) is an exact relation
satisfied by the full scaling function $H(z)$.
It is, however, still not easy to invert this double Laplace transform to extract
$H(z)$ explicitly from it. However, one can verify
that an ansatz 
\begin{equation}
H(z) = \frac{A}{z^{3/2}}\, e^{-B/z}\, ,
\label{ansatz_Hz.1}
\end{equation}
does satisfy Eq.~(\ref{dlt_scale.3}) with an appropriate choice for $A$ and $B$. Indeed,
let us fist substitute this ansatz on the LHS of (\ref{dlt_scale.3}) and perform the
integral over $z$ using the following identity
\begin{equation}
\int_0^{\infty} dz\, z^{\nu-1} \, e^{-B/z- a\, z}= 
2\, \left(\frac{B}{a}\right)^{\nu/2}\, K_{\nu}\left(
2\, \sqrt{a\, B}\right)\, \quad {\rm for}\quad a>0\,,\, B>0\, ,
\label{identity.2}
\end{equation}
where $K_{\nu}(z)$ is the modified Bessel function of the second kind. Using
$K_{1/2}(z)= \sqrt{\frac{\pi}{2z}}\, e^{-z}$ we get 
\begin{equation}
A \frac{\sqrt{\pi}}{\sqrt{B}}\, 
\frac{1}{\left(1+ 2\, \sqrt{B\, s}\right)}= \frac{1}{1+ \sqrt{s}}\, .
\label{match.1}
\end{equation}
This leads to the unique choices, $A= 1/\sqrt{4\pi}$ and $B=1/4$. Consequently we get the
result in Eq.~(\ref{Hz_scaling.1}).

\subsection{Matching between the two regimes}

\begin{figure}
\centering
\includegraphics[width=0.6\textwidth]{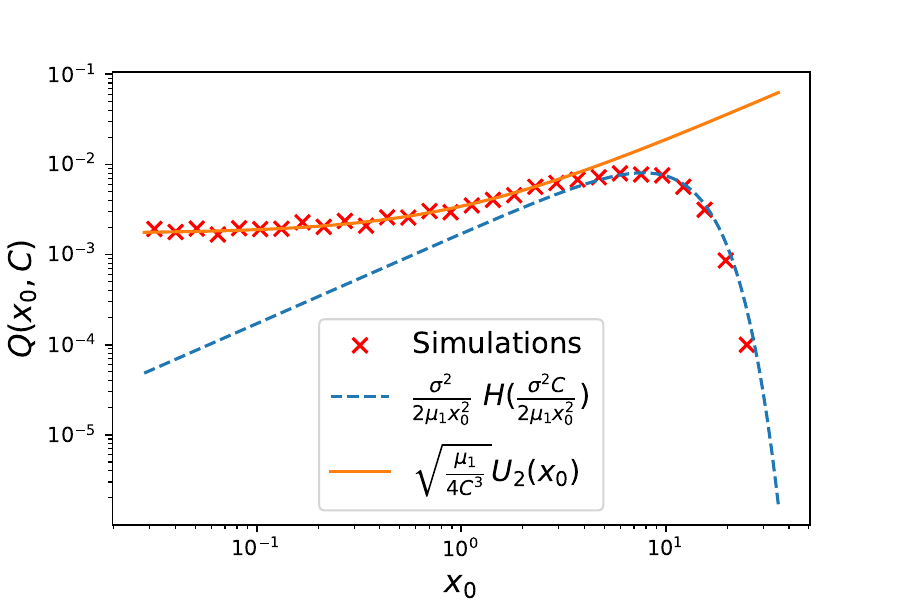}
\caption{Cost distribution $Q(x_0,C)$ as a function of $x_0$ for $C=30$, $f(\eta)=e^{-|\eta|}/2$, and $h(\eta)=|\eta|$. The lines correspond to the asymptotic behaviors in Eq.~\eqref{QXC_summary.1}. For this choice of the step distribution, we find $U_2(x_0)=(1+x_0)/\sqrt{\pi}$. The crosses are obtained from numerical simulations, averaging over $10^7$ trajectories.}
\label{fig:Q230}
\end{figure}

Let us first summarize the main results for $Q(x_0,C)$ for large $C$ in two different regimes 
of $x_0$, as derived
in the previous two subsections. We found that, as a function of $x_0$ for large $C$, the
following two regimes for the cost distribution
\begin{eqnarray}
\label{QXC_summary.1}
Q(x_0,C) \simeq \begin{cases}
&  \sqrt{\frac{\mu_1}{4}}\, \frac{1}{C^{3/2}}\, U_2(x_0) \hspace{3cm} {\rm for} \quad x_0\sim O(1) \\
\\
&  \frac{\sigma^2}{2 \, \mu_1\, x_0^2}\, 
H\left( \frac{\sigma^2\, C}{2\, \mu_1\, x_0^2}\right) \hspace{2.8cm} 
{\rm for} \quad x_0\sim O\left(\sqrt{C}\right) \, ,
\end{cases}
\end{eqnarray}
where $U_2(x_0)$ and $H(z)$ are given respectively in Eqs.~(\ref{Ux0_def}) and (\ref{Hz_scaling.1}). The crossover between the two regimes is shown in Fig.~\ref{fig:Q230}.

It is then interesting to check how these two regimes  match as $x_0$ increases from $O(1)$
to $O(\sqrt{C})$. Indeed, taking $x_0\gg 1$ in the first line of Eq.~(\ref{QXC_summary.1})
and using the large $x_0$ behavior of $U_2(x_0)$ from Eq.~(\ref{Ux0_asymp}), we get
\begin{equation}
Q(x_0,C) \simeq \sqrt{\frac{\mu_1}{2\,\pi\, \sigma^2}}\, \frac{x_0}{C^{3/2}}\, \quad {\rm for}
\quad 1\ll x_0\, .
\label{Qx0_asymp_lower.1}
\end{equation}
On the other hand, if we start in the regime $x_0\sim \sqrt{C}$ in the second line
of Eq.~(\ref{QXC_summary.1}) and take the limit $x_0\ll \sqrt{C}$, this corresponds to taking \red{the} 
$z\to \infty$ limit of the scaling function $H(z)$ where $H(z)\approx 1/\sqrt{4\pi z^3}$.
Using $z= \sigma^2 C/(2\mu_1 x_0^2)$, we then get
\begin{equation}
Q(x_0,C) \simeq \sqrt{\frac{\mu_1}{2\,\pi\, \sigma^2}}\, \frac{x_0}{C^{3/2}}\, \quad {\rm for}
\quad x_0\ll \sqrt{C}\, .
\label{Qx0_asymp_upper.1}
\end{equation}
Comparing Eqs.~(\ref{Qx0_asymp_lower.1}) and (\ref{Qx0_asymp_upper.1}), we see that $Q(x_0,C)$,
as a function of $x_0$ for large $C$, matches smoothly between the two regimes.

\section{L\'evy jumps with $h(\eta)=|\eta|$}
\label{section:levy}

In this section, we generalize our results to the L\'evy flights where the jump distribution $f(\eta)$ 
is symmetric, continuous and has
a power law tail, $f(\eta)\simeq D_\mu\, |\eta|^{-(\mu+1)}$ for large $|\eta|$ with the L\'evy index $0<\mu\le 2$.
This means that the Fourier transform of the jump distribution has the following behavior
\begin{equation}
\hat{f}(q)\simeq 1- \left(a_\mu\, |q|\right)^{\mu} \quad {\rm as} \quad q\to 0\, ,
\label{F_levy.1}
\end{equation}
where $a_\mu>0$ is the characteristic length of a L\'evy flight and is related to
the amplitude $D_\mu$ of the power law tail by the relation~\cite{BG_review}
\begin{equation}
D_\mu= a_\mu^{\mu}\, \frac{1}{\pi}\, \sin(\mu\pi/2)\, \Gamma(1+\mu)\, .
\label{Dmu_rel.1}
\end{equation}

For simplicity, we will choose the cost function per jump as $h(\eta)=|\eta|$. Hence, $C$ here describes the total length of the trajectory until the first-passage event. In this case, we need to distinguish between two cases: (i) when $1<\mu\le 2$ where the mean cost
per jump $\mu_1= \int_{-\infty}^{\infty} f(\eta)\, |\eta|\, d\eta$ is finite
and (ii) when $0<\mu\le 1$ where the mean cost per jump is divergent. Below, we consider
the two cases separately.

\subsection{The case $1<\mu\le 2$}

In this case, we distiguish between two regimes: (a) when the starting position $x_0\sim O(1)$
and (b) when $x_0\sim O\left(C^{1/\mu}\right)$.

\subsubsection{ When $x_0\sim O(1)$}

For $1<\mu\le 2$, the mean cost per jump $\mu_1$ is finite. Substituting $A(p)\simeq 1- \mu_1\, p \, $
for small $p$ from Eq.~(\ref{Ap_small.1}) in Eq.~(\ref{sol.2}) and setting $p=0$ in $\psi_1(\lambda, p)$ in (\ref{psi_def.1}) gives, as $p\to 0$,
\begin{equation}
\int_0^{\infty} dx_0\,  e^{-\lambda \, x_0}\, \int_0^{\infty} dC\,  Q(x_0,C)\,e^{-p\, C}\simeq
\frac{1}{\lambda}\, \left[1- \sqrt{\mu_1\, p}\,\, \psi_1(\lambda,0)\right]\, ,
\label{x1.11}
\end{equation}
where
\begin{equation}
\psi_1(\lambda,0)= \exp\left[-\frac{\lambda}{\pi}\,
\int_0^{\infty} \frac{\ln (1- \hat{f}(q))}{\lambda^2+q^2}\, dq\right]\, \quad
{\rm with}\quad \hat{f}(q)= \int_{-\infty}^{\infty} f(\eta)\, e^{i\, q\, \eta}\, d\eta\, .
\label{x1.21}
\end{equation}
The rest of the analysis is similar to Section (\ref{Sec:finite.1}), and we do not repeat here.
We get 
\begin{equation}
Q(x_0,C)\simeq \sqrt{\frac{\mu_1}{4}}\, \frac{1}{C^{3/2}}\, U_\mu(x_0)\, ,
\label{Qx0_scale.11}
\end{equation}
where the Laplace transform of the function $U_\mu(x_0)$ is given by 
\begin{equation}
\int_0^{\infty} dx_0\, e^{-\lambda\, x_0}\, U_\mu(x_0)= 
\frac{\psi_1(\lambda,0)}{\lambda\, \sqrt{\pi}}\, .
\label{Udef_12}
\end{equation}
The asymptotic behaviors of
the function $U_\mu(x_0)$ was analyzed in detail in Ref.~\cite{MMS_17} and for $1<\mu\le 2$, 
it reads
\begin{eqnarray}
\label{Umx0_asymp}
U_\mu(x_0)=\begin{cases}
& \frac{1}{\sqrt{\pi}}+ \alpha_1\, x_0 + O(x_0^2) \quad\quad\quad\quad\quad\quad 
{\rm as}\quad x_0\to 0 \\
\\
& A_\mu\, x_0^{\mu/2} + o\left(x_0^{\mu/2}\right)
\hspace{2cm} {\rm as} \quad x_0\to \infty\, ,
\end{cases}
\end{eqnarray}
where $\alpha_1$ is the same as in Eq.~(\ref{constat_def})
and the constant $A_\mu$ is given by
\begin{equation}
A_\mu= \frac{a_\mu^{-\mu/2}}{\sqrt{\pi}\, \Gamma(1+\mu/2)}\, .
\label{Amu_def}
\end{equation}
In this formula for $A_\mu$, the constant $a_\mu$
is the same that appears in the small $q$ expansion of the Fourier transform
of the jump distribution $f(\eta)$ in Eq.~(\ref{F_levy.1}).
Note that for $x_0=0$ and $1<\mu\le 2$, using $U_\mu(0)=1/\sqrt{\pi}$,
we get from Eq.~(\ref{Qx0_scale.11})
\begin{equation}
Q(0,C)\simeq \sqrt{\frac{\mu_1}{4\, \pi}}\, \frac{1}{C^{3/2}}\, ,
\label{Q0C.11}
\end{equation}
which matches with the general result derived below in Section \ref{subsec:asymptotic}.

\subsubsection{ When $x_0\sim O\left(C^{1/\mu}\right)$}

In this case we consider the scaling when $C\to \infty$ and $x_0\to \infty$, with the ratio
$C/x_0^{\mu}$ held fixed. Our main result here is to show that in this scaling limit,
$Q(x_0,C)$ exhibits the
scaling form
\begin{equation}
Q(x_0,C) \simeq \frac{1}{x_0^{\mu}}\, G_\mu\left( \frac{C}{x_0^{\mu}}\right)\, .
\label{Gmz_scaling.1}
\end{equation}
The scaling function $G_\mu(z)$ depends on the L\'evy index $\mu$ only,
but is otherwise universal, i.e., it does not depend on the details
of $f(\eta)$ as long as $\mu_1$ is finite.
For large $z$, the scaling function $G_\mu(z) \sim z^{-3/2}$ has a power law tail, while
it approaches a constant as $z\to 0$.

In this case, we need to analyze Eqs.~(\ref{sol.2}) and (\ref{psi_def.1})  in the
limit when $\lambda\to 0$ and $p\to 0$ with the ratio $\lambda/p^{1/\mu}=w$
held fixed.
To proceed, we first rescale $q=\lambda u$ in (\ref{psi_def.1})
which gives for $h(\eta)=|\eta|$,
\begin{equation}
\psi_1(\lambda, p)= \exp\left[- \frac{1}{\pi}\, \int_0^{\infty}
\frac{ \ln \left(1-\, \int_{-\infty}^{\infty} f(\eta)\,e^{-p\, |\eta|}\, e^{i\, \lambda\, u\,
\eta}\,
d\eta\right)}{1+u^2}\, du\right]\, .
\label{psi_def.21}
\end{equation}
Expanding the argument of the logarithm in small $\lambda$ and small
$p$, and using the definitions (\ref{mean_jump_def}) and (\ref{F_levy.1}), it is easy to see that, to leading order
for small $\lambda$ and small $p$ with $w= \lambda/p^{1/\mu}$ held fixed, one gets
\begin{equation}
1-\, \int_{-\infty}^{\infty} f(\eta)\,e^{-p\, |\eta|}\, e^{i\, \lambda\, u\,
\eta}\,
d\eta \simeq p\, \left(\mu_1+ a_\mu^{\mu}\, w^{\mu}\, u^{\mu}\right)\, .
\label{scale.11}
\end{equation}
Substituting this in Eq.~(\ref{psi_def.21}) and simplifying, we can then re-express Eq.~(\ref{sol.2})
in the scaling limit as
\begin{equation}
\int_0^{\infty} dx_0\, \int_0^{\infty} dC\,  Q(x_0,C)\,e^{-p\, C}\, e^{-\lambda \, x_0}
\simeq \frac{1}{\lambda}\left[1- J_\mu(w)\right]\, ,
\label{sol.21}
\end{equation}
where $J_\mu(w)$ is defined as
\begin{equation}
J_\mu(w)= \exp\left[- \frac{1}{\pi}\, \int_0^{\infty} \frac{du}{u^2+1}\, \ln\left(1+ 
(c_\mu\, w\, u)^{\mu}\right)\right]\, \quad {\rm where}\quad c_\mu= \frac{a_\mu}{(\mu_1)^{1/\mu}}\, .
\label{Jm_def}
\end{equation}
The same function $J_\mu(w)$ also appeared in Ref.~\cite{MMS_17} in the context of the survival probability
and its asymptotic behaviors were analyzed there. For instance, for $1<\mu\le 2$, the function
$J_\mu(w)$ has the asymptotic behaviors~\cite{MMS_17}
\begin{eqnarray}
\label{Jmw_asymp}
J_\mu(w) \simeq \begin{cases}
& 1- \alpha_\mu\, w- \beta_{\mu}\, w^{\mu}  \quad\quad\, {\rm as} \quad w\to 0 \\
\\
& \frac{1}{ (c_\mu w)^{\mu/2}}  \hspace{2.2cm} {\rm as} \quad w\to \infty\, ,
\end{cases}
\end{eqnarray}
where the constants $\alpha_\mu$ and $\beta_\mu$ are given by
\begin{equation}
\alpha_{\mu}= - \frac{c_\mu}{\sin(\pi/\mu)} \quad {\rm and}\quad 
\beta_{\mu}= \frac{c_\mu^{\mu}}{2 \cos(\mu \pi/2)}\, ,
\label{abmu_def}
\end{equation}  
and the constant $c_\mu$ is given in Eq.~(\ref{Jm_def}).

Now, substituting the anticipated scaling form (\ref{Gmz_scaling.1}) on the LHS
of (\ref{sol.21})  and making similar change of variables leading to (\ref{dlt_scale.2}) , 
we arrive at the `double Laplace transform'
\begin{equation}
\int_0^{\infty} dy\, e^{-y}\, \int_0^{\infty} dz~ G_\mu(z)\, e^{- y^{\mu} w^{-\mu}\, z}= 
1- J_\mu(w)\, .
\label{dlt_scale.21}
\end{equation}
We emphasize that this is an exact relation satisfied by the scaling function $G_\mu(z)$.
While it is difficult to invert this double Laplace transform to compute the full 
scaling function $G_\mu(z)$
explicitly, one can extract the asymptotic behaviors of $G_\mu(z)$ from Eq.~(\ref{dlt_scale.21})
by using the small and large $w$ behavior of $J_\mu(w)$ given in Eq.~(\ref{abmu_def}). Skipping details, we get
\begin{eqnarray}
\label{GMz_asymp}
G_\mu(z) \simeq \begin{cases}
& C_\mu \quad\,\, {\rm as} \quad z\to 0\\
\\
& \frac{B_\mu}{z^{3/2}}  \quad {\rm as} \quad z\to \infty\, ,
\end{cases}
\end{eqnarray}
where the constants $C_\mu$ and $B_\mu$ can be computed explicitly
\begin{equation}
C_\mu= \frac{c_\mu^{\mu} (\mu-1)}{2 \sin((\mu-1)\pi/2)\, \Gamma(2-\mu)}\, , \quad {\rm and}
\quad B_{\mu}= \frac{1}{ 2\sqrt{\pi} \Gamma(1+\mu/2)\, c_\mu^{\mu/2}}\, .
\label{CB_def}
\end{equation}

\subsubsection{Matching between the two regimes}

\begin{figure}
\centering
\includegraphics[width=0.6\textwidth]{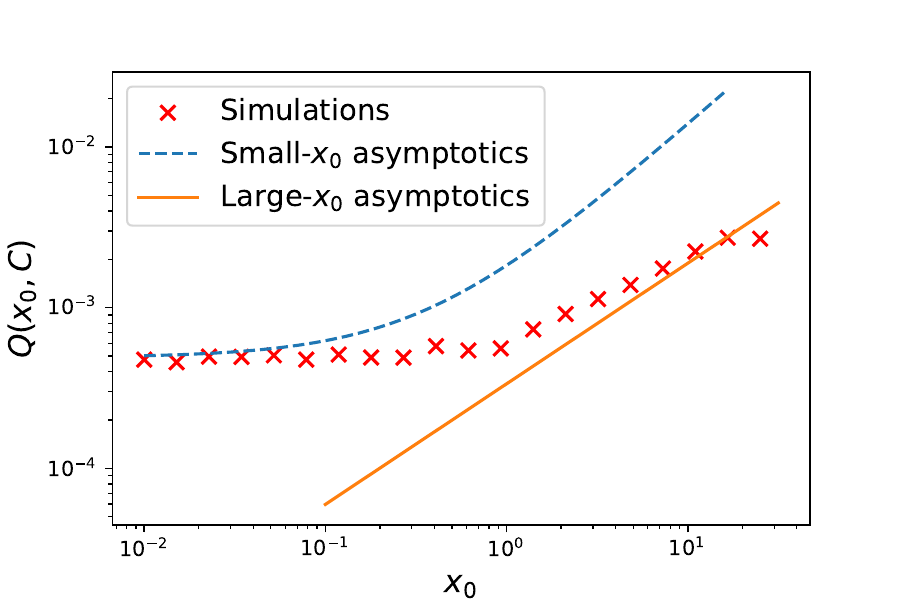}
\caption{Cost distribution $Q(x_0,C)$ as a function of $x_0$ for 
$C=100$, $f(\eta)=\mu /(2|\eta|)^{\mu}$ for $|\eta|>1$ with 
$\mu=3/2$, and $h(\eta)=|\eta|$. The lines correspond to the 
asymptotic behaviors in Eq.~\eqref{QXC_summary.11_explicit}. 
The crosses are obtained from numerical simulations, averaging over $5\times
10^6$ trajectories. The deviations from the theoretical line for large $x_0$ are due to finite size (finite $C$) effects.}
\label{fig:Qlevy15}
\end{figure}

Thus, summarizing for L\'evy flights with index $1<\mu\le 2$, we have the following behaviors
for the cost distribution, as a function of $x_0$ for large $C$,
following two regimes
\begin{eqnarray}
\label{QXC_summary.11}
Q(x_0,C) \simeq \begin{cases}
&  \sqrt{\frac{\mu_1}{4}}\, \frac{1}{C^{3/2}}\, U_\mu(x_0) \hspace{3cm} {\rm for} \quad x_0\sim O(1) \\
\\
& \frac{1}{x_0^{\mu}}\, G_\mu\left( \frac{C}{x_0^{\mu}}\right)\ \hspace{3.5cm}\,
{\rm for} \quad x_0\sim O\left(C^{1/\mu}\right) \, ,
\end{cases}
\end{eqnarray}
where the asymptotic behaviors of $U_\mu(x_0)$ and $G_\mu(z)$ are given respectively in (\ref{Umx0_asymp})
and (\ref{GMz_asymp}). Using the small-$x_0$ regime of $U_{\mu}(x_0)$ in Eq.~\eqref{Umx0_asymp} and the small-$z$ regime of $G_{\mu}(z)$ in Eq.~\eqref{GMz_asymp}, where $z=C/x_0^{\mu}$ here, we get the explicit asymptotic behaviors, valid for large $C$,
\begin{eqnarray}
\label{QXC_summary.11_explicit}
Q(x_0,C) \simeq \begin{cases}
  \sqrt{\frac{\mu_1}{4}}\, \frac{1}{C^{3/2}}\, \left[\frac{1}{\sqrt{\pi}}+\alpha_1x_0\right] \hspace{3cm} &{\rm for} \quad x_0\to 0 \\
\\ 
 \, \frac{B_{\mu}x_0^{\mu/2}}{ C^{3/2}}\ \hspace{3.5cm}\,
&{\rm for} \quad x_0\to\infty \, ,
\end{cases}
\end{eqnarray}
where $\alpha_1$ is given in Eq.~\eqref{constat_def} and $B_{\mu}$ in Eq.~\eqref{CB_def}. These asymptotic behaviors are shown in Fig.~\ref{fig:Qlevy15} and are in good agreement with numerical simulations.

To check that the two regimes in Eq.~\eqref{QXC_summary.11} match each other smoothly as $x_0$ increases from $O(1)$ to 
$O\left(C^{1/\mu}\right)$, we proceed as follows. Indeed, taking $x_0\gg 1$ in the first line
of Eq.~(\ref{QXC_summary.11}) and using the asymptotic large $x_0$ behavior of $U_\mu(x_0)$
from Eq.~(\ref{Umx0_asymp}), we get
\begin{equation}
Q(x_0,C)\simeq \sqrt{\frac{\mu_1}{4\pi}}\, \frac{a_\mu^{-\mu/2}}{\Gamma(1+\mu/2)}\,  
\frac{x_0^{\mu/2}}{C^{3/2}}\, , \quad {\rm for}\quad x_0\gg 1 \, .
\label{qxc_asymp1_12}
\end{equation}
In contrast, for $1\ll x_0\ll C^{1/\mu}$, we need to substitute the large $z$ behavior of $G_\mu(z)$
in the second line of Eq.~(\ref{QXC_summary.11}). Substituting (\ref{GMz_asymp}) and
using $c_\mu= a_\mu/(\mu_1)^{1/\mu}$ from Eq.~(\ref{Jm_def}),
we get
\begin{equation}
Q(x_0,C)\simeq 
\sqrt{\frac{\mu_1}{4\pi}}\, \frac{a_\mu^{-\mu/2}}{\Gamma(1+\mu/2)}\,
\frac{x_0^{\mu/2}}{C^{3/2}}\, , \quad {\rm for}\quad 1\ll x_0\ll C\ .
\label{qxc_asymp2_12}
\end{equation}
Clearly the two regimes match smoothly as $x_0$ increases.

\subsection{The case $0<\mu<1 $}

In this case, it turns out that there are again two principal
regimes of the cost distribution $Q(x_0,C)$ as a function of the starting position $x_0$
and fixed but large $C$: (a) when $x_0\sim O(1)$ and (b) when $x_0\sim O(C)$.

\subsubsection{ When $x_0\sim O(1)$}

Our starting point is again the central exact formula in 
Eqs.~(\ref{sol.2}) and (\ref{psi_def.1}), where $A(p)$ \red{in} Eq.~(\ref{sol.2}) is given by
\begin{equation}
A(p)= \int_0^{\infty} f(\eta)\, e^{-p\, \eta}\, d\eta\, ,
\label{AP_def01}
\end{equation}
Now, when $f(\eta)\simeq  D_\mu\, \eta^{-(\mu+1)}$ for large $\eta$ with $0<\mu<1$,
the two leading terms in the small $p$ expansion of $A(p)$ are given by
\begin{equation}
A(p) \simeq \frac{1}{2}- b_\mu\, p^{\mu}\, ,
\label{Ap_smallp.12}
\end{equation}
This is clearly different from the corresponding result for $A(p)=1-\mu_1\, p$ 
for the case $\mu>1$. The constant $b_\mu$ is related to the large $\eta$ amplitude $D_\mu$
by the relation  
\begin{equation}
D_\mu= b_\mu\, \frac{\mu}{\Gamma(1-\mu)}\, .
\label{Dmu_rel.2}
\end{equation}
Finally, eliminating $D_\mu$ from Eqs.~(\ref{Dmu_rel.1}) and (\ref{Dmu_rel.2}) provides
an exact relationship between the two constants $a_\mu$ and $b_\mu$ for $0<\mu<1$,
\begin{equation}
\frac{a_\mu^{\mu/2}}{\sqrt{b_\mu}}= \sqrt{2\, \cos(\mu \pi/2)}\, .
\label{amu_bmu_rel.1}
\end{equation}
We will see later that this relation plays a crucial role in establishing the exact matching of
the cost distribution function in two different scaling regimes. 
 
We now substitute this small $p$ behavior of $A(p)$ in 
Eq.~(\ref{Ap_smallp.12}) on the right hand side (RHS)
of Eq.~(\ref{sol.2}). Since we are interested in the regime $x_0\sim (1)$, or equivalently
$\lambda$ fixed, we can further set $p=0$ in $\Psi_1(\lambda, p)$ on the RHS
of (\ref{sol.2}) leading to
\begin{equation}
\int_0^{\infty} dx_0\,  e^{-\lambda \, x_0}\, \int_0^{\infty} dC\,  Q(x_0,C)\,e^{-p\, C}\simeq
\frac{1}{\lambda}\, \left[1- \sqrt{2\, b_\mu\, p^{\mu}}\, \psi_1(\lambda,0)\right]\, ,
\label{x1.01}
\end{equation}
where $\psi_1(\lambda,0)$ is given in Eq.~(\ref{x1.21}).
We next take a derivative of Eq.~(\ref{x1.01}) with respect to $p$, and then invert
the Laplace transform with respect to $p$ using the identity
\begin{equation}
{\cal L}_{p\to C}^{-1}\left[ \frac{1}{p^{1-\mu/2}}\right]= \frac{1}{\Gamma(1-\mu/2)}\, 
\frac{1}{C^{\mu/2}}\, .
\label{linv_p01}
\end{equation}
This gives
\begin{equation}
\int_0^{\infty} dx_0\, e^{-\lambda\, x_0}\, C\, Q(x_0,C) \simeq 
\sqrt{\frac{b_\mu\,\pi}{2}}\,
\frac{\mu}{\Gamma(1-\mu/2)}\, \frac{1}{C^{\mu/2}}\, 
\frac{\psi_1(\lambda,0)}{\lambda\, \sqrt{\pi}}\, .
\label{linv2_p01}
\end{equation}
This implies that
\begin{equation}
Q(x_0,C) \simeq \sqrt{\frac{b_\mu\,\pi}{2} } \,
\frac{\mu}{\Gamma(1-\mu/2)}\, \frac{1}{C^{\mu/2+1}}\, U_\mu(x_0)\, ,
\label{qxc1_01}
\end{equation}
where the Laplace transform of $U_\mu(x_0)$ is given by
\begin{equation}
\int_0^{\infty} dx_0\, e^{-\lambda\, x_0}\, U_\mu(x_0)=\frac{\psi_1(\lambda,0)}{\lambda\, \sqrt{\pi}}\, . 
\label{Ux_def_01}
\end{equation}
This is identical to the case $1<\mu<2$ in Eq.~(\ref{Udef_12})
and indeed the asymptotic behaviors of the function $U_\mu(x_0)$ detailed in
Eq.~(\ref{Umx0_asymp}) also hold for $0<\mu<1$.

Note that for $x_0=0$, using $U_\mu(0)= 1/\sqrt{\pi}$, we get from Eq.~(\ref{qxc1_01}) that for large $C$
\begin{equation}
Q(0,C) \simeq \sqrt{\frac{b_\mu}{2}}\,
\frac{\mu}{\Gamma(1-\mu/2)}\, \frac{1}{C^{\mu/2+1}}\, .
\label{q0c1_01}
\end{equation}
In contrast, for large $x_0$,
using $U_\mu(x_0)\simeq A_\mu\, x_0^{\mu/2}$ from Eq.~({\ref{Umx0_asymp}) we get, for large $C$,
\begin{equation}
Q(x_0,C) \simeq \left[ \sqrt{2b_\mu}\, a_\mu^{-\mu/2}\, \frac{\sin(\mu\pi/2)}{\pi} \right]\,
\frac{ x_0^{\mu/2}}{C^{\mu/2+1}}\, ,
\label{qxc2_01}
\end{equation}
where we used the expression for $A_\mu$ in Eq.~(\ref{Amu_def}).

\subsubsection{ When $x_0\sim O(C)$}

Unlike in the $1<\mu\le 2$ case, it turns out that for $0<\mu<1$, the scaling
regime occurs when $x_0\sim O(C)$. In other words, the scaling regime $x_0\sim C^{1/\mu}$
valid for $1<\mu\le 2$ freezes to $x_0\sim O(C)$ when $\mu$ decreases below $1$.
For $x_0\sim C$, it is natural to make the scaling ansatz
\begin{equation}
Q(x_0,C) \simeq \frac{1}{x_0}\, G_\mu\left( \frac{C}{x_0}\right) \quad {\rm for}\quad x_0\sim C\, ,
\label{qxc_scaling_01}
\end{equation}
where the scaling function $G_\mu(z)$ (with $0<\mu<1$) needs to be extracted from Eqs. 
(\ref{sol.2}) and (\ref{psi_def.1}).
Before extracting the scaling function, it is useful to note that for the choice $h(\eta)=|\eta|$, the
total cost $C$ can not be less than $x_0$, and hence the scaled variable $z=C/x_0\ge 1$. In other words
the scaling function $G_\mu(z)$ is supported over the interval $z\in [1,\infty]$.

To proceed, we substitute the ansatz (\ref{qxc_scaling_01}) on the LHS of (\ref{sol.2}) and get
\begin{eqnarray}
\label{LHS_01}
\int_0^{\infty} dx_0\, \int_0^{\infty} dC\,  Q(x_0,C)\,e^{-p\, C}\, e^{-\lambda \, x_0}
\simeq \int_0^{\infty} dx_0\, e^{-\lambda\, x_0}\int_1^{\infty} dz\, G_\mu(z)\, e^{-p\, x_0\, z} = \int_1^{\infty} dz \frac{G_\mu(z)}{\lambda+p z}\, ,
\end{eqnarray}
where we made the change of variable $C=x_0\, z$ and \red{used} the fact that $z\ge 1$, in going from the first to the second line.
Since we want to take the limit $C\to \infty$, \red{$x_0\to\infty$} with the ratio $C/x_0=z$ fixed, this amounts to
taking the limit $p\to 0$ and $\lambda\to 0$, with the \red{ratio} $w=\lambda/p$ fixed. In this scaling limit,
Eq.~(\ref{LHS_01}) reads
\begin{equation}
\int_0^{\infty} dx_0\, \int_0^{\infty} dC\,  Q(x_0,C)\,e^{-p\, C}\, e^{-\lambda \, x_0}  \simeq \frac{1}{\lambda} \, \int_1^{\infty} dz\, \frac{G_\mu(z)}{1+ \frac{z}{w}}\, .
\label{LHS1_01}
\end{equation}
We now consider the RHS of Eq.~(\ref{sol.2}) in the same scaling limit. Substituting the small $p$
expansion of $A(p)$ from Eq.~(\ref{Ap_smallp.12}) into the RHS of (\ref{sol.2}) and equating both sides
(after canceling the factor $1/\lambda$ from both sides), we arrive at the relation
\begin{equation}
\int_1^{\infty} dz\, \frac{G_\mu(z)}{1+ \frac{z}{w}}= 1- 
\lim_{p\to 0}\left[ \sqrt{2\, b_\mu\, p^\mu}\, \psi_1(wp, p)\right]\, .
\label{scale_rel1_01}
\end{equation}
Note that the scaling function $G_\mu(z)$ is normalized to unity, i.e., $\int_1^{\infty} G_\mu(z)\, dz=1$.
Subtracting $1$ from both sides and using the normalization of $G_\mu(z)$ gives a simplified relation
\begin{equation}
\int_1^{\infty} dz\, G_\mu(z)\, \frac{z}{z+w}= \lim_{p\to 0}\left[ \sqrt{2\, b_\mu\, p^\mu}\, \psi_1(wp, p)\right]\, .
\label{scale_rel2_01}
\end{equation}

Thus, we now need to determine the limiting value on the RHS of Eq.~(\ref{scale_rel2_01}) where
$\psi_1(wp,p)$ is defined in Eq.~(\ref{psi_def.1}) and reads, upon rescaling $q=\lambda \, u$,
\begin{equation}
\psi_1(wp,p) = \exp\left[- \frac{1}{\pi}\int_0^{\infty} \frac{du}{1+u^2}\, \ln \left(1- \int_{-\infty}^{\infty}
f(\eta) e^{-p |\eta|+ i\, p\, w\, u\, \eta}\, d\eta \right)\right]\, .
\label{psi_lim1_01}
\end{equation} 
To take the small $p$ limit, we proceed as follows. Noting that $f(\eta)$ is a symmetric function of $\eta$,
we can replace $e^{ i\,p\, w\,u\,\eta}$ by $\cos(p\,w\,u\,\eta)$ on the RHS of (\ref{psi_lim1_01}). Let us then rewrite
\begin{eqnarray}
\label{psi_lim2_01}
\int_{-\infty}^{\infty} f(\eta)\, e^{-p |\eta|} \cos(p\,w\, u\,\eta)\, d\eta &=&
\int_{-\infty}^{\infty} f(\eta) e^{-p|\eta|}\, d\eta - \int_{-\infty}^{\infty} f(\eta)\, e^{-p |\eta|}\,
\left[1- \cos(p\,w\,u\,\eta)\right]\, d\eta \nonumber \\
&=& 2 A(p) - \frac{2}{p} \int_0^{\infty} f\left(\frac{y}{p}\right)\, e^{-y}\, \left[1- \cos(w\,u\, y)\right]\, dy\, ,
\end{eqnarray}
where we recall that $A(p)=\int_0^{\infty} f(\eta)\, e^{-p \eta}\, d\eta$ and in going from the second to the
third \red{equation} we made a change of variable $p\eta=y$ and used the symmetry of the integrand explaining
the factor $2$. We are now ready to take the limit $p\to 0$. The first term $2 A(p)\simeq 1- 2\, b_\mu\, p^{\mu}$
as $p\to 0$ from Eq.~(\ref{Ap_smallp.12}). In the second term, as $p\to 0$, the argument of $f(y/p)$ becomes large
and we can replace it by its power-law tail $f(y/p)\simeq D_\mu \red{(y/p)}^{-(1+\mu)}$, provided the
integral remains convergent (which it does). Hence to leading order for small $p$, we get
\begin{equation}
\int_{-\infty}^{\infty} f(\eta)\, e^{-p |\eta|} \cos(p\,w\, u\,\eta)\, d\eta \simeq
1- 2\, b_\mu\, p^{\mu} - 2\, D_\mu\, p^{\mu} \, \int_0^{\infty} \frac{dy}{y^{\mu+1}}\, e^{-y}\, \left[1-\cos( w\, u\, y)\right]\, dy\, .
\label{psi_lim3_01}
\end{equation}
Using further the relation between $D_\mu$ and $b_\mu$ in Eq.~(\ref{Dmu_rel.2}) and simplifying we get
\begin{equation}
1- \int_{-\infty}^{\infty} f(\eta)\, e^{-p |\eta|} \cos(p\,w\, u\,\eta)\, d\eta \simeq 2\, b_\mu\, p^{\mu}\,\left[1+
\frac{\mu}{\Gamma(1-\mu)}\, I_\mu(wu)\right]\, ,
\label{psi_lim4_01}
\end{equation}
where the function $I_\mu(x)$ is given by
\begin{equation}
I_\mu(x)= \int_0^{\infty} \frac{dy}{y^{\mu+1}}\, e^{-y}\, \left[1- \cos(x\,y)\right]\, .
\label{Ix_def.01}
\end{equation} 
Substituting (\ref{psi_lim4_01}) on the RHS of (\ref{psi_lim1_01}) one gets in the limit $p\to 0$ with $w$ fixed
\begin{equation}
\psi_1(wp,p)\simeq \frac{1}{\sqrt{2\, b_\mu\, p^{\mu}}}\, \exp\left[- \frac{1}{\pi}\, \int_0^{\infty}
\frac{du}{1+u^2}\, \ln\left( 1+ \frac{\mu}{\Gamma(1-\mu)}\, I_\mu(w\, u)\right) \right]\, .
\label{psi_lim5_01}
\end{equation}
Consequently, we can take the limit on the RHS  of Eq.~(\ref{scale_rel2_01}) and this leads to an 
exact relation satisfied by our scaling function $G_\mu(z)$
\begin{equation}
\int_1^{\infty} dz\, G_\mu(z)\, \frac{z}{z+w}= \exp\left[- \frac{1}{\pi}\,
\int_0^{\infty} \frac{du}{1+u^2}\, \ln \left[1+ \frac{\mu}{\Gamma(1-\mu)}\,
I_\mu(w\,u)\right] \right]\, ,
\label{relation.1}
\end{equation}
where $I_\mu(x)$ is defined in Eq.~(\ref{Ix_def.01}).

While this relation (\ref{relation.1}) is exact for all $0<\mu<1$, it is still difficult to extract the full scaling function
$G_\mu(z)$ from this relation. However, as we show now the large $z$ asymptotic behavior of $G_{\mu}(z)$
can be extracted from Eq.~(\ref{relation.1}). To extract the large $z$ behavior of $G_\mu(z)$, we need
to investigate the large $w$ behavior of the RHS of (\ref{relation.1}). For this, we need to know how
$I_\mu(x)$ in Eq.~(\ref{Ix_def.01}) behaves for large $x$. Making the change of variable $xy=v$ in
(\ref{Ix_def.01}), we get for large $x$
\begin{equation}
I_\mu(x) \simeq x^\mu\, \int_0^{\infty} \frac{dv}{v^{\mu+1}} (1-\cos v)= 
\frac{\Gamma(1-\mu)}{\mu}\, \cos(\mu\pi/2)\, x^{\mu}\, .
\label{Ix_largex_01}
\end{equation}
For large $w$ on the RHS of (\ref{relation.1}), we can then substitute this large $x$ behavior of $I_\mu(x)$
and perform the resulting integrals. In particular, we used the identity
\begin{equation}
\int_0^{\infty} \frac{du}{1+u^2}\, \ln (u) =0\, .
\label{integral_idem.01}
\end{equation}
Simplifying, we get the large $w$ behavior of the RHS of (\ref{relation.1}) as
\begin{equation}
\int_1^{\infty} dz\, G_\mu(z)\, \frac{z}{z+w}\simeq \frac{1}{\sqrt{\cos(\mu\pi/2)}}\, \frac{1}{w^{\mu/2}}\, .
\label{rhs_largew.01}
\end{equation}
Now, we need to make a guess on how $G_\mu(z)$ behaves for large $z$ so that the LHS of (\ref{rhs_largew.01})
behaves as $w^{-\mu/2}$ for large $w$. It is not difficult to see that indeed $G_\mu(z)$ decays as a
power law
\begin{equation}
G_\mu(z)\simeq \frac{B_\mu}{z^{1+\mu/2}}  \quad {\rm as}\quad z\to \infty\, .
\label{Gmu_largez.01}
\end{equation}
We now rescale $z=w u$ on the LHS of (\ref{rhs_largew.01}) and take the large $w$ limit 
assuming (\ref{Gmu_largez.01}). This gives the LHS
\begin{equation}
\int_1^{\infty} dz\, G_\mu(z)\, \frac{z}{z+w} \simeq 
\frac{B_\mu}{w^{\mu/2}}\, \int_0^{\infty} \frac{du}{u^{\mu/2}\, (u+1)} = 
\frac{B_\mu}{w^{\mu/2}}\, \frac{\pi}{\sin(\mu\pi/2)}\, .
\label{lhs1.01}
\end{equation}
Comparing to the RHS in Eq.~(\ref{rhs_largew.01}) shows that the ansatz (\ref{Gmu_largez.01}) is indeed correct
and moreover, the amplitude $B_\mu$ has the expression
\begin{equation}
B_\mu= \frac{1}{\pi}\, \frac{\sin(\mu\pi/2)}{\sqrt{\cos(\mu\pi/2)}}\, .
\label{Bmu_def}
\end{equation}

\subsubsection{Matching between the two regimes}

\begin{figure}
\centering
\includegraphics[width=0.6\textwidth]{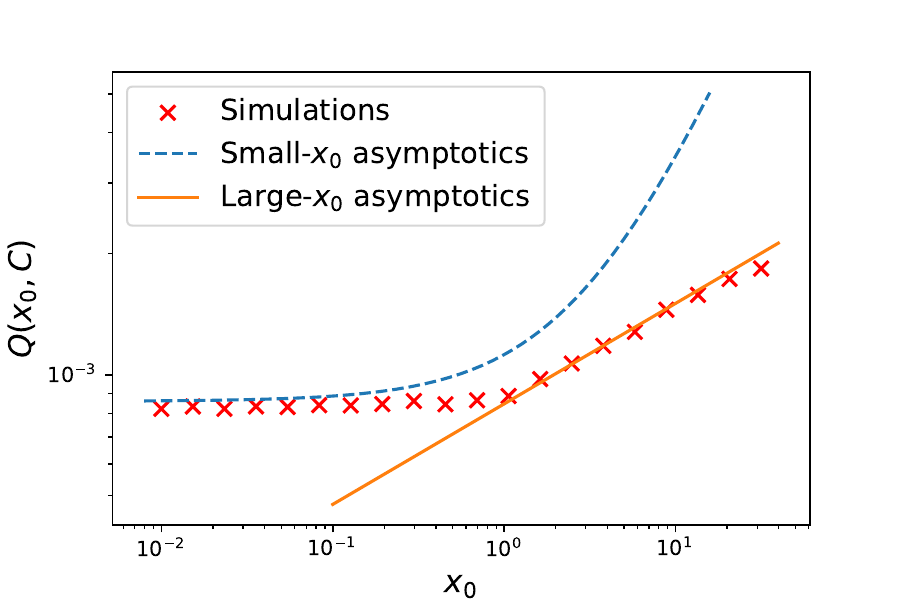}
\caption{Cost distribution $Q(x_0,C)$ as a function of $x_0$ for 
$C=100$, $f(\eta)=\mu /(2|\eta|^{\mu})$ for $|\eta|>1$ with $\mu=1/2$, 
and $h(\eta)=|\eta|$. The lines correspond to the asymptotic behaviors 
in Eq.~\eqref{QxC_summary.01_explicit}. The crosses are obtained from 
numerical simulations, averaging over $5\times 10^6$ trajectories. The 
deviations from the theoretical line for large $x_0$ are due to finite 
size (finite $C$) effects.}
\label{fig:Qlevy05}
\end{figure}

Let us then summarize first our main results for the cost distribution $Q(x_0,C)$ for the case $0<\mu<1$.
For large but fixed $C$, as a function of $x_0$, the distribution $Q(x_0,C)$ has the 
following two regimes
\begin{eqnarray}
\label{QxC_summary.01}
Q(x_0,C) \simeq \begin{cases}
&  \sqrt{\frac{b_\mu \pi}{2}}\, \frac{\mu}{\Gamma(1-\mu/2)}
\frac{1}{C^{\mu/2+1}}\,
U_\mu(x_0) \hspace{3cm} {\rm for} \quad x_0\sim O(1) \\
\\
& \frac{1}{x_0}\, G_\mu\left( \frac{C}{x_0}\right)\ \hspace{5.5cm}
{\rm for} \quad x_0\sim O(C) \, ,
\end{cases}
\end{eqnarray}
where the asymptotic behavior of $U_\mu(x_0)$ is
given in Eq.~(\ref{Umx0_asymp}) even for $0<\mu<1$. The scaling function
\red{$G_\mu(z)$} is supported over $z\in [0,1]$ and satisfies an exact transform relation (\ref{relation.1}).
For large $z$, the scaling function $G_\mu(z)$ decays as a power law
\begin{equation}
G_\mu(z)\simeq \frac{B_\mu}{z^{1+\mu/2}} \quad {\rm where}\quad B_\mu= \frac{1}{\pi}\, 
\frac{\sin(\mu\pi/2)}{\sqrt{\cos(\mu\pi/2)}} \, .
\label{Gmuz_largez1}
\end{equation}
Therefore, using Eq.~\eqref{Umx0_asymp} and \eqref{Gmuz_largez1}, we find the explicit asymptotic behaviors, valid for large $C$,
\begin{eqnarray}
\label{QxC_summary.01_explicit}
Q(x_0,C) \simeq \begin{cases}
  \sqrt{\frac{b_\mu \pi}{2}}\, \frac{\mu}{\Gamma(1-\mu/2)}
\frac{1}{C^{\mu/2+1}}\,
\left(\frac{1}{\sqrt{\pi}}+ \alpha_1\, x_0\right) \hspace{3cm} &{\rm for} \quad x_0\to 0  \\
\\
B_{\mu} \frac{x_0^{\mu/2}}{C^{1+\mu/2}}\ \hspace{5.5cm}
& {\rm for} \quad x_0\to \infty\, ,
\end{cases}
\end{eqnarray}
where $\alpha_1$ is given in Eq.~\eqref{constat_def}. These asymptotic results are shown in Fig.~\ref{fig:Qlevy05} and are in good agreement with numerical simulations.

It would be important to check how these two regimes match as $x_0$ increases from $O(1)$ to $O(C)$ for large
$C$. Indeed, taking $x_0\gg 1$ in the first line of (\ref{QxC_summary.01}), we get the result
in Eq.~(\ref{qxc2_01}), namely
\begin{equation}
Q(x_0,C) \simeq \left[ \sqrt{2b_\mu}\, a_\mu^{-\mu/2}\, \frac{\sin(\mu\pi/2)}{\pi} \right]\,
\frac{ x_0^{\mu/2}}{C^{\mu/2+1}}\, , \quad {\rm for}\quad x_0\gg 1\, .
\label{qxc2_03}
\end{equation}
In contrast, for $1\ll x_0\ll C$, we need to substitute the large $z$ behavior of $G_\mu(z)$
in the second line of Eq.~(\ref{QxC_summary.01}). Substituting (\ref{Gmuz_largez1}) we get 
\begin{equation}
Q(x_0,C)\simeq \frac{1}{\pi}\, \frac{\sin(\mu\pi/2)}{\sqrt{ \cos(\mu\pi/2) }}\, 
\frac{x_0^{\mu/2}}{C^{\mu/2+1}}\, , \quad {\rm for}\quad 1\ll x_0\ll C\ .
\label{qxc2_04}
\end{equation}
In order that these two limiting behaviors in Eqs.~(\ref{qxc2_03}) and (\ref{qxc2_04}) match with each other, we
need to prove the identity
\begin{equation}
\left[ \sqrt{2b_\mu}\, a_\mu^{-\mu/2}\, \frac{\sin(\mu\pi/2)}{\pi} \right]=\frac{1}{\pi}\,
\frac{\sin(\mu\pi/2)}{\sqrt{\cos(\mu\pi/2)}}\, .
\label{iden_final.1}
\end{equation}
However, this is precisely guaranteed by the exact relation (\ref{amu_bmu_rel.1}). This proves
that for large $C$, the cost distribution in the two regimes of $x_0$ match smoothly 
with each other.

\section{The case $x_0=0$: Explicit results for $Q(C)$}

\label{sec:x0zero}

In this section, we focus on the case $x_0=0$. Let us recall the general formula for $Q(C)\equiv Q(0,C)$, derived in Section \ref{Model},
\begin{equation}
\int_0^{\infty} dC\, Q(C)\, e^{-p\, C}=1-\sqrt{1- 2\, A(p)}\,,\quad \text{ where }\quad A(p)= \int_0^{\infty} f(\eta)\, e^{-p\, h(\eta)}\, d\eta\,.
\label{formula_def_2}
\end{equation}
We recall that this general result was first obtained in Refs.~\cite{ACEK14,BCP22,Pozzoli_phd_thesis}. Below, we consider a number of models in which the Laplace transform in \eqref{formula_def_2} can be inverted exactly.

\subsection{Run-and-tumble particle model}
\label{subsec:RTP}

For the RTP case, where $h(\eta)=|\eta|$ and $f(\eta)$ is given in Eq.~\eqref{exp_jump.1}, $Q(C)$
is precisely the distribution of the total energy spent by
the particle till its first return to its starting point. For the double-exponential jump distribution in Eq.~\eqref{exp_jump.1} and the cost function $h(\eta)=|\eta|$ corresponding to energy being consumed at a constant rate, the Laplace transform \eqref{formula_def_2} can be explicitly inverted to yield
the distribution of the cost of
first return to the origin
\begin{equation}
Q(C)= \frac{1}{2}\, e^{-C/2}\left[ I_0\left(\frac{C}{2}\right)
- I_1\left(\frac{C}{2}\right)\right]\, ,
\label{sol_exp_final1}
\end{equation}
where $I_0(z)$ and $I_1(z)$ are modified Bessel functions (see Fig. \ref{fig:Q_exp}). Note that, since $h(\eta)=|\eta|$, the variable $C$ also describes the total distance traveled by the particle until first passage to the origin.

\begin{figure}
    \centering
    \includegraphics[width=0.6\textwidth]{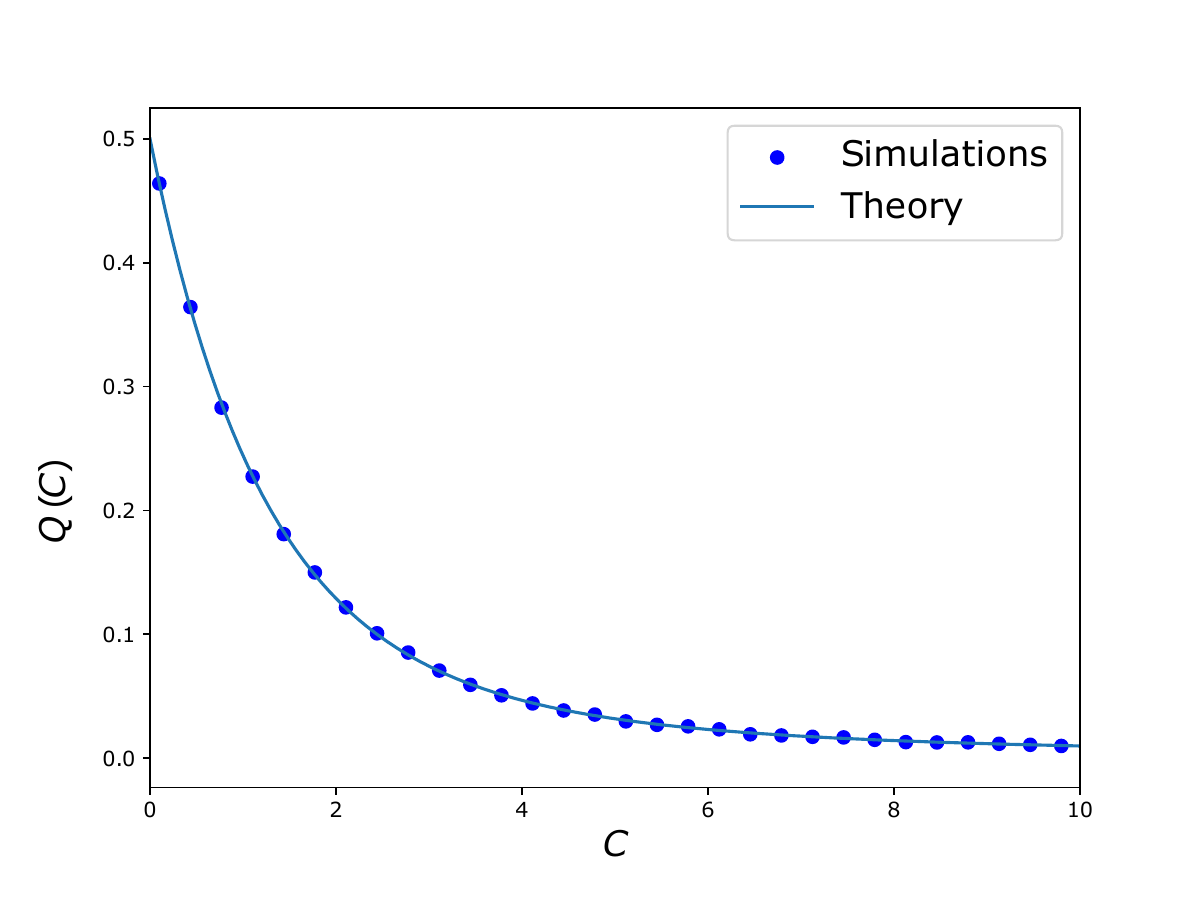}
    \caption{Distribution $Q(C)$ of the cost $C$ until first return for jump distribution $f(\eta)=e^{-|\eta|}/2$ and cost function $h(\eta)=|\eta|$. The continuous blue line corresponds to the exact result in Eq.~\eqref{sol_exp_final1} while the dots are obtained from numerical simulations with $10^6$ samples.}
    \label{fig:Q_exp}
\end{figure}

The
distribution has asymptotic behaviors
\begin{eqnarray}
Q(C) \approx \begin{cases}
&\frac{1}{2}- \frac{3}{8}\, C+ \frac{5}{32}\, C^2 \quad {\rm as}\quad C\to 0
\\
\\
& \frac{1}{2\sqrt{\pi}}\, \frac{1}{C^{3/2}} \quad\quad\quad\quad\, {\rm as}
\quad C\to \infty\ .
\end{cases}
\label{exp_asymp}
\end{eqnarray}
The asymptotic power law decay $\sim C^{-3/2}$ for large cost is universal for sufficiently light-tailed jump distributions (see Eq.~\eqref{Q0C.1}), and can be
understood by a heuristic scaling argument. The distribution $P(n_f)$ of the first-passage
time $n_f$ for random walks with symmetric and continuous jump distributions
is universal and given by the Sparre Andersen law~\cite{SA_54}
(see also Refs.~\cite{Feller, Redner_book, Persistence_review, fp_book_2014}).
For large $n_f$, we have $P(n_f)\approx (1/2\sqrt\pi)n_f^{-3/2}$. Now, since the cost $C$ in $n_f$ steps scale typically as $C\sim n_f$, it
follows that the distribution of $C$ for large $C$ also has exactly
the same power law tail. The fact that also the prefactors match is just a coincidence in this special case. In general, the prefactor is different and nontrivial and
we compute it exactly later.

\subsection{ Gaussian steps}
Let us 
consider the case when
\begin{equation}
f(\eta)= \frac{1}{\sqrt{\pi}}\, e^{-\eta^2}\, , \quad \,\, {\rm and}\quad
h(\eta) =\eta^2\ .
\label{sp_case.2}
\end{equation}
This corresponds to a random walk with Gaussian jump distribution
and a quadratic cost function, which is also a rather natural choice.
In this case, the exact formula \eqref{formula_def_2} gives
\begin{equation}
\tilde{Q}_p(0)= \int_0^{\infty} Q(C)\, e^{-p\, C}\, dC= 1-
\sqrt{1- \frac{1}{\sqrt{1+p}}}\ .
\label{sp2_lap.1}
\end{equation}
One needs some tricks to invert this Laplace transform. Let us first
define 
\begin{equation}
Q(C)= e^{-C}\, G(C)\ .
\label{G_def}
\end{equation}
Substituting \eqref{G_def} in \eqref{sp2_lap.1} and denoting $s=p+1$, we get
\begin{equation}
\int_0^{\infty} G(C)\, e^{-s\, C}\, dC= 1-\sqrt{1- \frac{1}{\sqrt{s}}}\ .
\label{G_lap.1}
\end{equation}
We now expand the r.h.s. of \eqref{G_lap.1} in a power series in $1/\sqrt{s}$
and invert each term by using the identity
\begin{equation}
{\cal L}^{-1}_{s\to C}\left[ s^{-n/2}\right]= 
\frac{C^{n/2-1}}{\Gamma(n/2)}\ .
\label{G_inv.1}
\end{equation}
This leads to a power series for $G(C)$ 
\begin{equation}
G(C)= \frac{2}{C}\, \sum_{n=1}^{\infty} \frac{\Gamma(2n-1)}{\Gamma(n)\,
\Gamma(n+1)\, \Gamma(n/2)\, 4^n}\, C^{n/2}\ .
\label{power_series_G}
\end{equation}
Using Mathematica, we resum this series and express
it in terms of hypergeometric functions. Finally, using the relation
\eqref{G_def}, we get a nice and explicit expression for $Q(0,C)$
\begin{equation}
Q(0,C)= \frac{e^{-C}}{2\sqrt{\pi\, C}}\, _2F_2\left[1/4,\, 
3/4; \, 1/2,\, 3/2;C\right] + \frac{e^{-C}}{8}\,
_2F_2\left[3/4,\, 
5/4; \, 3/2,\, 2;C\right]\ .
\label{exact_sp.2}
\end{equation}
One can easily plot this function (see Fig.~\ref{fig:Q_gauss}) and it has the asymptotic behaviors
\begin{align}
Q(0,C)\approx \begin{cases}
& \frac{1}{2\sqrt{\pi\, C}} \quad\quad\,\,\,\, {\rm as}\quad C\to 0 \\
\\
& \frac{1}{2\, \sqrt{2\,\pi}\, C^{3/2}} \quad {\rm as}\quad C\to \infty \ .
\end{cases}
\label{Gaussian_asymp}
\end{align}
The large $C$ decay $C^{-3/2}$ is in accordance with the general result discussed in Section \ref{subsec:asymptotic} below, while it also diverges as $C\to 0$. Comparing to the result for the
RTP case in Eq.~\eqref{exp_asymp}, we indeed see that the small $C$
behavior is rather nonuniversal, while the large $C$ result is a universal $C^{-3/2}$ law, as long as the mean cost per jump $\mu_1$ is finite.

\begin{figure}
    \centering
    \includegraphics[width=0.6\textwidth]{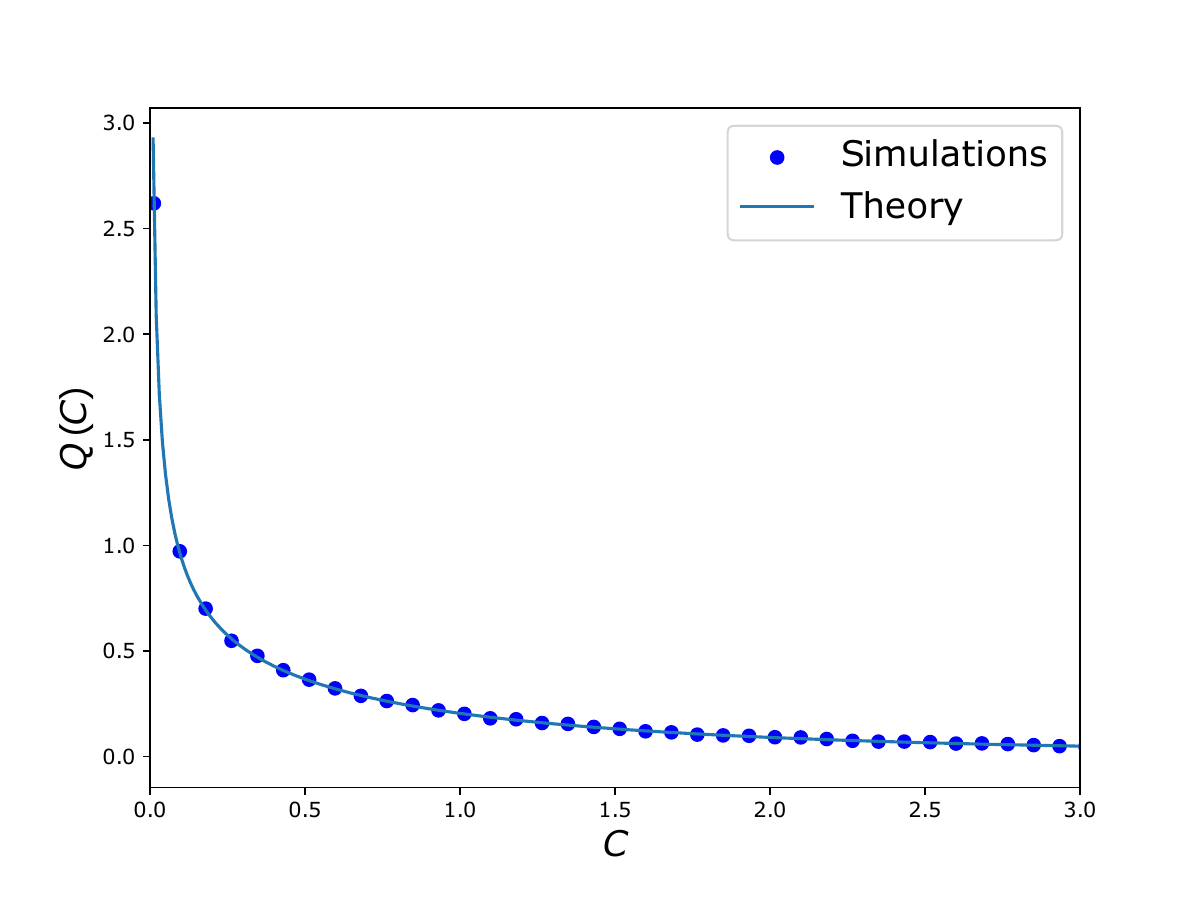}
    \caption{Distribution $Q(C)$ of the cost $C$ until first return for jump distribution $f(\eta)=e^{-\eta^2}/\sqrt{\pi}$ and cost function $h(\eta)=\eta^2$. The continuous blue line corresponds to the exact result in Eq.~\eqref{exact_sp.2} while the dots are obtained from numerical simulations with $10^6$ samples.}
    \label{fig:Q_gauss}
\end{figure}

\subsection{L\'evy flights}

We consider the following model
\begin{equation}
f(\eta)= \frac{1}{2 \, \sqrt{4\, \pi\, |\eta|^3}}\, e^{- 1/(4 |\eta|)}\, ; \quad\quad {\rm and}\quad h(\eta)=|\eta|\ .
\label{ex3.2}
\end{equation}
This is thus an example of a L\'evy flight where the jump distribution 
$f(\eta)\sim 1/|\eta|^{\mu+1}$ for large $|\eta|$, with a L\'evy exponent $\mu=1/2$.
Consider the following identity
\begin{equation}
\int_0^{\infty} \frac{a}{4 \pi t^{3}}\, e^{-a^2/{4t}}\, e^{- p\,t} dt= e^{-a\, \sqrt{p}}\ ,
\label{iden3.1} 
\end{equation}
valid for all $a>0$. Setting $t=\eta$ and $a=1$ in this identity, we then get for
the choice in Eq.~\eqref{ex3.2}
\begin{equation}
A(p)= 2 \int_0^{\infty} e^{-p h(\eta)}\, f(\eta)\, d\eta= e^{-\sqrt{p}}\, .
\label{ex3.Ap}
\end{equation}
Hence, Eq.~\eqref{formula_def_2} gives
\begin{equation}
\tilde{Q}_p(0)= \int_0^{\infty} e^{-p\, C}\, Q(C)\, dC= 1-\sqrt{1-e^{-\sqrt{p}}}\ .
\label{ex3_exact}
\end{equation}
We now use the following power series expansion
\begin{equation}
(1-x)^{1/2}= 1 - \sum_{m=1}^{\infty} \frac{(2m-2)!}{m!\, (m-1)!}\, \frac{x^m}{2^{2m-1}}\ ,
\label{ex3_power1}
\end{equation}
to write Eq.~\eqref{ex3_exact} as
\begin{equation}
\tilde{Q}_p(0)= 
\sum_{m=1}^{\infty} \frac{(2m-2)!}{m!\, (m-1)!\, 2^{2m-1}}\, e^{-m\, \sqrt{p}}\ .
\label{ex3_power2}
\end{equation}
We can now invert this Laplace transform with respect to $p$ term by term using the same identity
as in Eq.~\eqref{iden3.1} by setting $a=m$ for the inversion of the $m$-th term. This then
gives an explicit result
\begin{equation}
Q(C)= \frac{1}{\sqrt{4\pi C^3}}\, \sum_{m=1}^{\infty} 
 \frac{(2m-2)!}{(m-1)!\, (m-1)!\, 2^{2m-1}}\, e^{-m^2/{4C}}\ .
\label{ex3_explicit}
\end{equation}
We could not resum this series, but \red{we} can easily evaluate it
numerically and plot it as a function of $C$, as shown in Fig.~\ref{fig:Q_levy}.
The asymptotic behavior of $Q(0,C)$ for small and large $C$ can be easily derived
from Eq.~\eqref{ex3_explicit}. For example, for small $C$, the $m=1$ term dominates
giving, $Q(0,C) \simeq e^{-1/(4C)}/\sqrt{16 \pi C^3}$. On the other hand, for large $C$
the dominant contribution to the sum in Eq.~\eqref{ex3_explicit} comes from large $m$,
where we can use Stirling's formula to approximate the factorials and then the sum can be
converted to an integral which can be explicitly evaluated. Summarizing, we get
the following asymptotic behaviors
\begin{align}
Q(0,C) \approx \begin{cases}
&\frac{1}{2\, \sqrt{4\,\pi\, C^3} }\, e^{- 1/(4C)} \quad\quad {\rm as}\quad C\to 0 \\
\\
& \frac{1}{4 \Gamma(3/4)}\, \frac{1}{ C^{5/4}}\, \quad\quad\quad\quad\, {\rm as}
\quad C\to \infty\ .
\end{cases}
\label{exp3_asymp}
\end{align}
The large $C$ power law decay $C^{-5/4}$ is consistent with the general result $\mu/2+1$
for $\mu= 1/2<1$, as derived below in Section \ref{subsec:asymptotic}.

\begin{figure}
\includegraphics[width=0.6\textwidth]{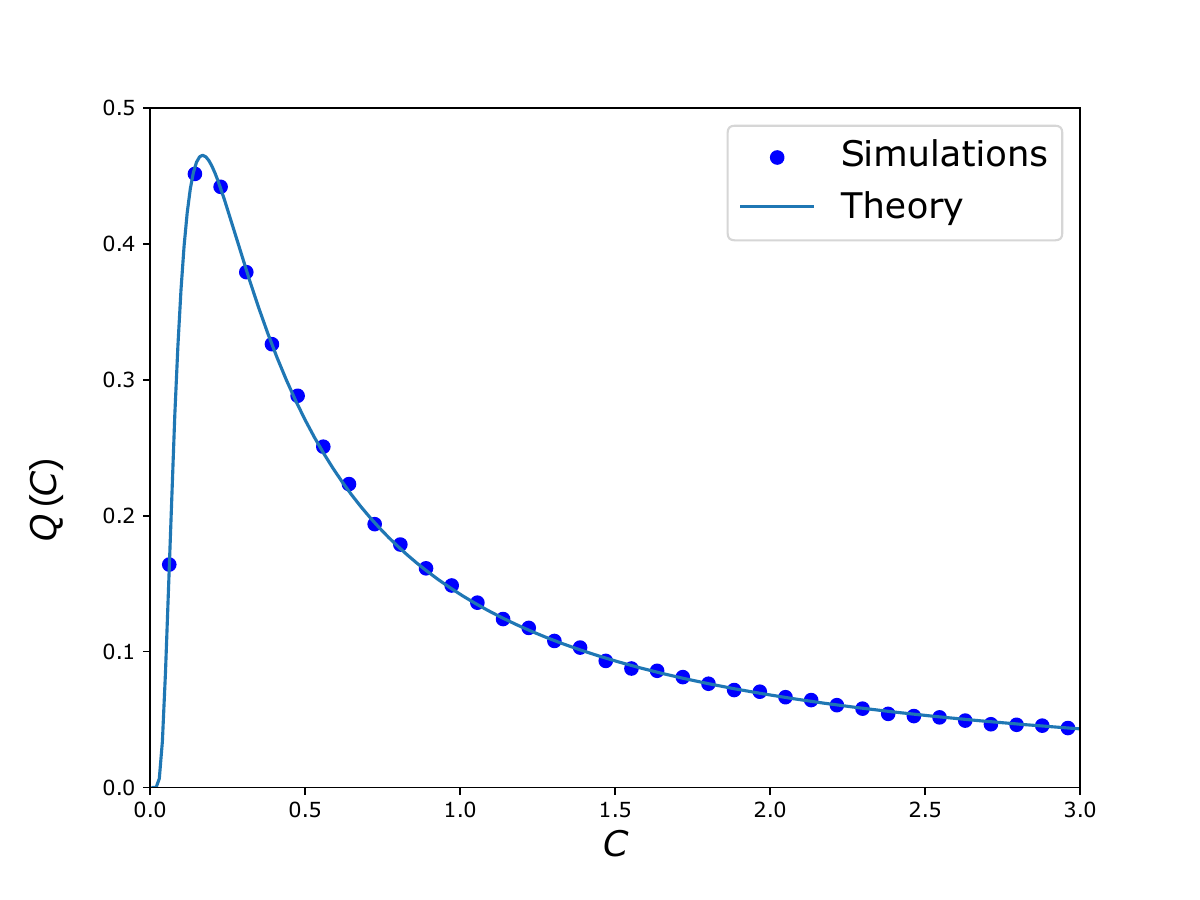}
\caption{Distribution $Q(C)$ of the cost $C$ until first return for jump distribution $f(\eta)= \frac{1}{2 \, \sqrt{4\, \pi\, |\eta|^3}}\, e^{- 1/(4 |\eta|)}$ and cost function $h(\eta)=|\eta|$. The continuous blue line corresponds to the exact result in Eq.~\eqref{ex3_explicit} while the dots are obtained from numerical simulations with $10^7$ samples.}
\label{fig:Q_levy}
\end{figure}

\subsection{Number of steps longer than a threshold}

Let us consider the cost function $h(\eta)=\theta(|\eta|-\eta_c)$, where $\theta(z)$ is the Heaviside step function and $\eta_c\geq 0$ is fixed. Using this cost function, the variable $C$ is the number of steps of length greater than $\eta_c$ until the first passage event. Note that when $\eta_c=0$, we have $q=0$ and $C$ reduces to the first passage time. We find
\begin{equation}
    A(p) = \frac12 \left[q +e^{-p}(1-q)\right]\,,
\end{equation}
where 
\begin{equation}
    q=\int_{-\eta_c}^{\eta_c}f(\eta)d\eta\,,
\end{equation}
is the probability that a single step is shorter than $\eta_c$. Plugging this expression for $A(p)$ and inverting the Laplace transform, we get
\begin{equation}
    Q(0)=1-\sqrt{1-q}\,,
    \label{eq:Q0}
\end{equation}
and 
\begin{equation}
    Q(C)=\sqrt{1-q}~\frac{(2(C-1))!}{C!(C-1)!}2^{-2C+1}
    \label{eq:QC}
\end{equation}
for $C\geq 1$. This result in Eqs.~\eqref{eq:Q0} and \eqref{eq:QC} is in perfect agreement with numerical simulations (see Fig.~\ref{fig:steps}).

Interestingly, for $C\geq 1$, the probability of $C$ can be written as
\begin{equation}
    Q(C)=\sqrt{1-q}~F_C\,,
\end{equation}
where $F_C(0)$ is the first passage probability to the origin after $C$ steps \red{(see Eq.~\eqref{eq:fn0_intro}).}. From Eq.~\eqref{eq:Q0}, we immediately find that $\sqrt{1-q}$ is the probability that there is at least \red{one} step of length greater than $\eta_c$ up to the first passage time. As a consequence, we find that the probability of $C$ conditioned on the event $C\geq 1$ is exactly the same as the probability distribution of the first passage time
\begin{equation}
    Q(C|C\geq 1)=F_C(0)\,.
\end{equation}
Remarkably, this last result is completely universal, as $F_C(0)$ is independent of the jump distribution $f(\eta)$.

\begin{figure}
\centering
\includegraphics[width=0.6\textwidth]{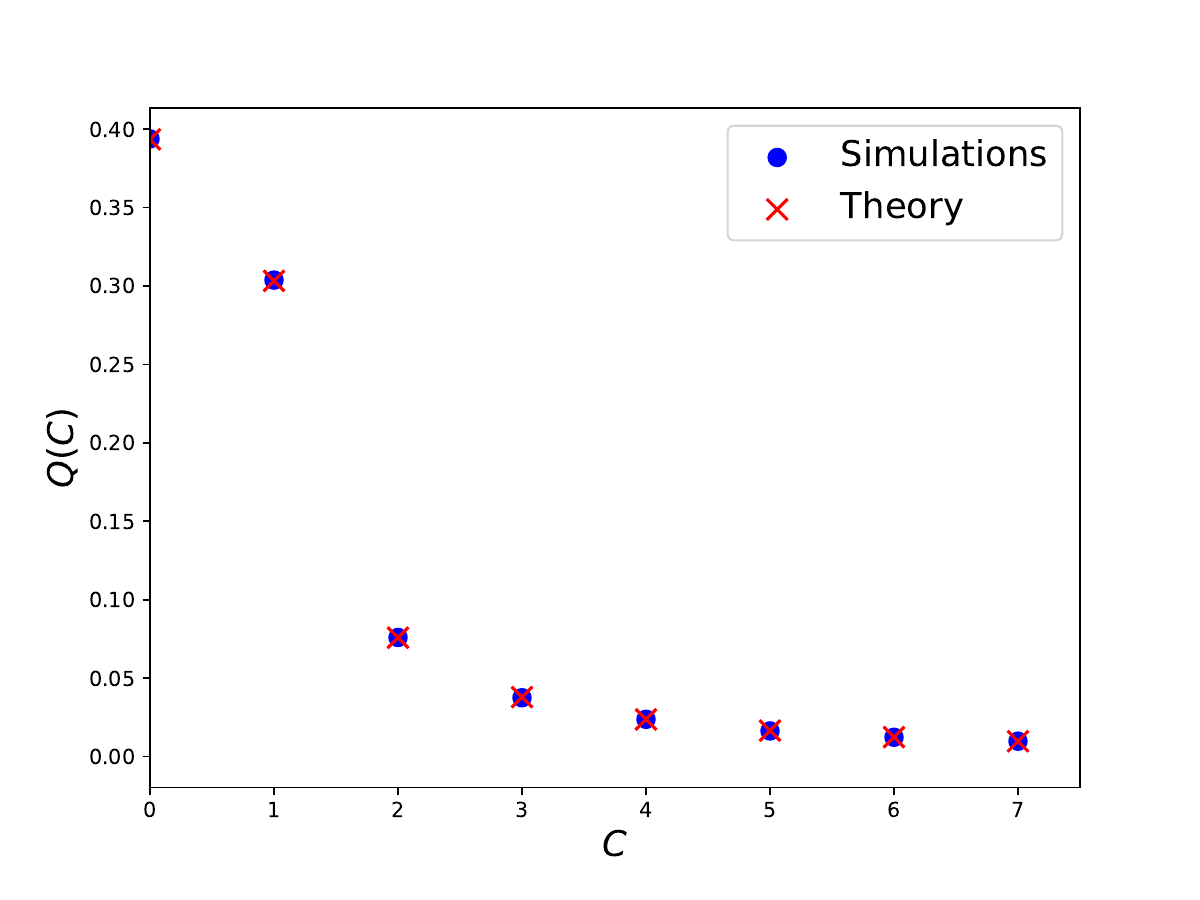}
\caption{ Probability distribution of the number $C$ of steps of length $|\eta|$ larger than $\eta_c=1$. The simulations are obtained from $10^6$ samples with step function $f(\eta)=e^{-|\eta|}/2$.}
\label{fig:steps}
\end{figure}

\subsection{Asymptotics}
\label{subsec:asymptotic}

Let us first consider the small $p$ limit
of Eq.~\eqref{formula_def_2}. By expanding in powers of $p$, one finds
that to leading order in small $p$
\begin{equation}
\tilde{Q}_p(0)= 1- a\, \sqrt{p} + \mathcal{O}(p)\ , \quad a=
\sqrt{2\, \int_0^{\infty} h(\eta)\, f(\eta)\, d\eta}\, .
\label{smallp.1}
\end{equation}
This is quite generic as long as the constant $a=\sqrt{\mu_1}$
above exists, where $\mu_1=\int_{-\infty}^{\infty} h(\eta)\, f(\eta)\, d\eta$
is just the mean cost per jump. In that case, the small $p$ expansion
in \eqref{smallp.1} indicates, by the Tauberian theorem, that 
$Q(0,C)$ has a universal power law tail for large $C$,
\begin{equation}
Q(0,C) \simeq \frac{a}{2\,\sqrt{\pi}\, C^{3/2}} \quad {\rm as}\,\, C\to \infty\ ,
\label{QC_tail.1}
\end{equation}
as previously derived in Eq.~\eqref{Q0C.1}.

If the mean cost per jump $\mu_1=\int_{-\infty}^{\infty} 
h(\eta)\, f(\eta)\, d\eta$ is divergent then the analysis above
does not hold. Consider, for instance, the case where
$h(\eta)=|\eta|$ (linear cost) and 
the jump distribution has a power law tail $f(\eta)\sim 1/|\eta|^{\mu+1}$
for large $|\eta|$ with the exponent $0<\mu\le 2$ corresponding
to L\'evy flights. In this case, clearly $\mu_1$ is divergent
for $0<\mu<1$ and convergent for $\mu>1$. For $\mu>1$, one will
again find the $C^{-3/2}$ decay of the cost distribution for large $C$. However, for
$0<\mu<1$, one finds that
\begin{equation} 
A(p)= \int_0^{\infty} e^{-p\, h(\eta)}\, f(\eta)\, d\eta 
\sim \frac{1}{2}- b_\mu\, p^{\mu} \quad {\rm as}\quad p\to 0\ ,
\label{Levy.1}
\end{equation}
where $b_\mu$ is an unimportant positive constant.
Consequently, from Eq.~\eqref{formula_def_2}, we find that to 
leading order for small $p$
\begin{equation}
\tilde{Q}_p(0) \approx 1- \sqrt{2\, b_\mu}\,\, p^{\mu/2}\ .
\label{Levy.2}
\end{equation}
This indicates, again via the Tauberian theorem, that for large $C$,
the cost distribution has a power law tail: $Q(0,C)\sim C^{-(1+\mu/2)}$, in agreement with the result in Eq.~\eqref{q0c1_01}. Thus, summarizing for the L\'evy jump distributions with L\'evy index
$0<\mu\le 2$, the cost distibution $Q(0,C) \sim C^{-\theta}$ for large
$C$ where the exponent $\theta$ is given by
\begin{align}
\theta = \begin{cases}
& \frac{\mu}{2}+1 \quad {\rm for}\quad 0<\mu<1 \\
\\
& \frac{3}{2}  \quad\quad\quad {\rm for} \quad 1< \mu\le 2\ .
\end{cases}
\label{Levy.3}
\end{align}
Thus the exponent $\theta$ increases linearly with $\mu$ for $\mu\in [0,1]$
and then freezes at the value $3/2$ for $\mu>1$, with a logarithmic
correction exactly at $\mu=1$ (for example for the Cauchy jump distribution $f(\eta) = 1/[\pi (\eta^2+1)]$). 

It turns out that unlike the universal behavior of $Q(0,C)$ for large $C$
as discussed above, the small $C$ behavior is rather non-universal
and it depends on the choices of $f(\eta)$ and $h(\eta)$. But even these non-universal behaviors can be extracted 
from the exact result in \eqref{formula_def_2}.

\section{Conclusions} 
\label{sec:conclusions}

We have derived a general formula in \blue{Eqs.~\eqref{sol.2} and 
(\ref{psi_def.1})}
for the 
distribution of the total cost $C$ incurred by a one-dimensional random 
walker until its first-passage to the origin valid for arbitrary starting 
position $x_0\geq 0$, provided the jump distribution $f(\eta)$ and the 
cost function $h(\eta)>0$ are symmetric and continuous. 
\blue{Our main focus in this paper was on how the distribution
$Q(x_0,C)$ behaves in  different regimes of $x_0$.
For both random walks and L\'evy flights, we have 
analyzed in detail the different scaling regimes that emerge in the 
limit of large $C$ and large $x_0$, finding several universal 
scaling functions that only depend on the L\'evy index $0<\mu\le 2$}.

\blue{In this paper, we have restricted ourselves only to a single cost variable
$C$ associated with the random walk $x_m$. However,
our result naturally generalizes to the case of multiple
cost variables
$C^i=\sum_{m=1}^{n_f}h_i(\eta_m)$ with $i=1,2,\ldots, N$, each
with its own $h_i(\eta)$ as in Eq. (\ref{cost.1_intro}), \red{which} are correlated 
through the noise term of the 
basic random walk in Eq.~\eqref{markov.1}.} For instance, in the case $x_0=0$, 
the $N$-fold Laplace transform of the joint distribution 
$Q(0,C^1,\ldots,C^N)$ of the cost variables until the
first return of the random walk to the origin satisfies
\begin{align}
 \Big\langle e^{-\sum_{i=1}^N p_i C^i}\Big\rangle = 1- \sqrt{1-2 A(p_1,p_2,\ldots, p_N)}\label{eq:extensions}
\end{align}
where 
\begin{equation}
A(p_1,p_2,\ldots, p_N)= \int_0^{\infty} e^{-\sum_{i=1}^N p_i h_i(\eta)} f(\eta) d\eta\ .
\end{equation}
A thorough analysis of the consequences of the generalized 
formula \eqref{eq:extensions} is deferred to a future publication.

It would be interesting to consider extensions of our setting to higher 
dimensions and different velocity distributions for the RTP model. 
Nonlinear cost functions are also particularly interesting to 
investigate in this context \cite{MMV23.1,MMV23.2}, as well as modified 
cost processes subject to an independent noise term as well. 
Additionally, it would be interesting to extend our framework to 
resetting processes.

{\bf Acknowledgments:} We thank R. Artuso and G. Pozzoli for useful discussions and for pointing out relevant references. This work was supported by a Leverhulme Trust International Professorship grant [number LIP-202-014]. SNM  acknowledges support from ANR Grant No. ANR-23-CE30-0020-01 EDIPS. P.V. acknowledges support from UKRI Future Leaders Fellowship Scheme (No. MR/S03174X/1). For the purpose of open access, the authors have applied a creative commons attribution (CC BY) licence to any author-accepted manuscript version arising.

\appendix

\section{Explicit solution of the backward integral equation for a special case}
\label{appendix}

In this section, we derive an exact solution to the integral equation \eqref{FT_bfp.2}. First of all, we consider the case of additive costs, i.e., $p(\eta, \xi) = f(\eta)\, \delta(\xi-h(\eta))$, so that Eq.~\eqref{FT_bfp.2} reduces to
\begin{align}
\tilde{Q}_p(x_0) = \int_0^{\infty} dx_1\, \tilde{Q}_p(x_1)\, 
e^{-p\, h(x_1-x_0)}\, f(x_1-x_0)+ \int_{-\infty}^{-x_0} e^{-p\, h(\eta_1)}\, f(\eta_1)\, d\eta_1\,.
\label{back_int.2s}
\end{align}
To proceed, we consider the following step distribution and cost function
\begin{equation}
f(\eta)= \frac{1}{2}\, e^{-|\eta|} \, ; \quad \,\, {\rm and}\quad 
h(\eta)=|\eta| \,,
\label{sp_case.1}
\end{equation}
corresponding to the RTP model in Section \ref{subsec:RTP}. Substituting the choice (\ref{sp_case.1}) in the integral equation
(\ref{back_int.2s}) one gets
\begin{equation}
\tilde{Q}_p(x_0)= \frac{1}{2}\, 
\int_0^{\infty} dx_1\, \tilde{Q}_p(x_1)\, e^{-(p+1)\, |x_1-x_0|}
+ \frac{1}{2}\, \int_{x_0}^{\infty} e^{-(p+1)\, \eta}\, d\eta \, ,
\label{exp_int.1}
\end{equation}
where we used the symmetry to change the limits of the integral
in the second term of the r.h.s in Eq.~\eqref{back_int.2s}. 
This integral equation can be reduced to a differential equation
following the trick used in numerous contexts before, see e.g.
Ref.~\cite{CM_2005,MCZ_06} (including an application to black hole
physics as in the recent paper~\cite{MMB_24}). This trick uses the following identity
\begin{equation}
\frac{d^2}{dx^2}\left[ e^{-a\, |x-b|}\right]= -2\, a\, \delta(x-b)
+ a^2\, e^{-a\, |x-b|}\ .
\label{iden.1}
\end{equation}

Differentiating Eq.~\eqref{exp_int.1} twice with respect to $x_0$
and using the identity \eqref{iden.1} simply gives, for $x_0\ge 0$,
\begin{equation}
\frac{d^2 \tilde{Q}_p(x_0)}{dx_0^2}= p(p+1)\, \tilde{Q}_p(x_0)\ .
\label{diff_int.1}
\end{equation}
The general solution, for $x_0\ge 0$, is trivially
\begin{equation}
\tilde{Q}_p(x_0)= A\, e^{-\sqrt{p(p+1)}\, x_0} + B\, e^{\sqrt{p(p+1)}\, x_0}\ ,
\label{sol_exp.1}
\end{equation}
where $A$ and $B$ are arbitrary constants. First we notice that when
$x_0\to \infty$, the solution $\tilde{Q}_p(x_0)$ can not diverge exponentially. This immediately fixes $B=0$. However, finding the constant $A$
is more tricky since we do not have any available boundary condition
on $\tilde{Q}_p(x_0)$. The important point is that in arriving
at the differential equation from the integral equation, we
took two derivatives and thus we lost some information, including
constant and linear terms in $x_0$. Hence, we need to ensure that the 
solution of the differential
equation $\tilde{Q}_p(x_0)= A\, e^{-\sqrt{p(p+1)}\, x_0}$ actually satisfies also 
the integral equation \eqref{diff_int.1}. Indeed, substituting back this solution
into the integral equation \eqref{diff_int.1} we find that this is
indeed the solution of the integral equation as well, provided
\begin{equation}
A= 1- \sqrt{\frac{p}{p+1}}\ .
\label{A.1}
\end{equation}
This then fixes the solution uniquely as
\begin{equation}
\tilde{Q}_p(x_0)= \left[1- \sqrt{\frac{p}{p+1}}\right]\, 
e^{-\sqrt{p(p+1)}\, x_0}\, .
\label{sol_exp.2}
\end{equation}

Setting, in particular, $x_0=0$ for simplicity, we get
\begin{equation}
\tilde{Q}_p(0)= \int_0^{\infty} Q (0,C)\, e^{-p\, C}\, dC=
1- \sqrt{\frac{p}{p+1}}\ .
\label{sol_exp.3}
\end{equation}
It turns out that a direct inversion of this Laplace transform
is slightly difficult. To circumvent this problem,
we first differentiate both sides with respect to $p$ to obtain
\begin{equation}
\int_0^{\infty} e^{-p\, C}\, C\, Q(0,C)\, dC= \frac{1}{2 \sqrt{p} (1+p)^{3/2}}\ .
\label{sol_exp.4}
\end{equation}
Now this Laplace transform can be inverted using Mathematica and we 
explicitly obtain the distribution of the cost of
first return to the origin 
\begin{equation}
Q(0,C)= \frac{1}{2}\, e^{-C/2}\left[ I_0\left(\frac{C}{2}\right)
- I_1\left(\frac{C}{2}\right)\right]\ ,
\label{sol_exp_final}
\end{equation}
where $I_0(z)$ and $I_1(z)$ are modified Bessel functions. This result coincides with the one in Eq.~\eqref{sol_exp_final1}.

\end{document}